%% file: main.tex
\pdfoutput=1

\documentclass[10pt, conference,romanappendices,letterpaper]{IEEEtran}

\usepackage[utf8]{inputenc}

\input{control-wasiur}

\setboolean{longVersion}{true}

\IEEEoverridecommandlockouts

\title{Collaborative Uploading in Heterogeneous Networks: Optimal  and Adaptive Strategies\ifthenelse{\boolean{longVersion}}{\\(Extended Version)}{}
}

\author{\IEEEauthorblockN{Wasiur R. KhudaBukhsh\IEEEauthorrefmark{1}, Bastian Alt\IEEEauthorrefmark{1}, Sounak Kar\IEEEauthorrefmark{2}, Amr Rizk\IEEEauthorrefmark{2},  and Heinz Koeppl\IEEEauthorrefmark{1}}
	\IEEEauthorblockA{\IEEEauthorrefmark{1}Bioinspired Communication Systems Lab (BCS),  \{wasiur.khudabukhsh \textbar~bastian.alt \textbar~heinz.koeppl\}@bcs.tu-darmstadt.de,}
	\IEEEauthorblockA{\IEEEauthorrefmark{2}Multimedia Communications Lab (KOM),  \{sounak.kar \textbar~amr.rizk\}@kom.tu-darmstadt.de, \\
		Technische Universit\"at Darmstadt, Germany}
}

\IEEEoverridecommandlockouts

\begin{document}
	\alglanguage{pseudocode}
\maketitle


\IEEEpeerreviewmaketitle

\begin{abstract}
\input{sections/0_abstract}

\end{abstract}

\input{sections/1_introduction}

\input{sections/Model}
\input{sections/Streaming_problem}

\input{sections/Numerical}

\input{sections/Related_Work}
\input{sections/Discussion.tex}
\appendices
\input{sections/Appendix}

\ifthenelse{\boolean{longVersion}}{\input{sections/Appendix_B} }

\ifthenelse{\boolean{longVersion}}{\input{sections/Appendix_C} }

\section*{Acknowledgment}
This work has been funded by the German Research Foundation~(DFG) as part of projects C3 and B4 within the Collaborative Research Center~(CRC) 1053 -- MAKI. Computational facilities provided by the Lichtenberg - High Performance Computer at TU Darmstadt are  gratefully acknowledged.
%

\bibliographystyle{IEEEtran}
\input{main.bbl}


\end{document}

%% file: control-wasiur.tex
\usepackage{graphicx}
\usepackage{subcaption}
\usepackage{cleveref}
\usepackage{amssymb}
\usepackage{mathtools}
\usepackage{algorithm}
\usepackage{algpseudocode}
\usepackage{algpascal}
\usepackage{color}
\usepackage{enumitem}

\usepackage{amsmath}
\usepackage[cmintegrals]{newtxmath}
\usepackage{bm} 

\graphicspath{ {Figures/} }
\usepackage{subcaption}
\usepackage[T1]{fontenc}
\usepackage{cite}
\usepackage{dsfont}

\makeatletter
\algnewcommand\Input{\item[\textbf{Input:}]}%
\algnewcommand\Output{\item[\textbf{Output:}]}%
\algnewcommand\Init{\item[\textbf{Init:}]}%
\makeatother

\usepackage{amstext} 
\usepackage{array}   
\newcolumntype{L}{>{$}l<{$}} 
\newcolumntype{C}{>{$}c<{$}} 
\newcolumntype{R}{>{$}r<{$}} 

\usepackage{amsthm}
\theoremstyle{plain}

\theoremstyle{definition}
\newtheorem{myRemark}{Remark}
\newtheorem{myDefinition}{Definition}
\newtheorem{myExample}{Example}

\newcommand{\setOfNaturals}{\mathbb{N}}
\newcommand{\setOfNonnegativeIntegers}{\mathbb{N}_0}

\newcommand{\setN}[1]{ [ #1]}

\newcommand{\effectiveDomain}[1]{\mathcal{D}(#1)}

\newcommand{\NegBin}[2]{\text{NB}(#1, #2)}

\newcommand{\cardinality}[1]{\mid #1 \mid}
\newcommand{\myDoperator}[2]{\mathrm{M}_{#1}\,#2}
\newcommand{\myMuoperator}[2]{\mathrm{\mu}_{#1 }\,#2}

\DeclarePairedDelimiter\floor{\lfloor}{\rfloor}

\renewcommand{\vec}[1]{\mathbf{#1}}

\newcommand{\differential}[1]{\, \mathrm{d} #1}

\DeclareMathOperator*{\argmin}{arg\,min}

\newcommand{\indicator}[1]{\mathds{1}(#1)}

\newcommand{\myExp}[1]{\exp \bigl( #1 \bigr)  }

\newcommand{\Permanent}[1]{\text{per} \hspace{2pt} #1}
\newcommand{\allpermutations}[1]{\Theta (#1)}
\newcommand{\diophantine}[2]{\Lambda (#1,#2)}
\newcommand{\genAllocation}[3]{\Upsilon (#1,#2, #3)}

\newcommand{\projection}[2]{\mathsf{P}_{#1}{\left( #2 \right)}}

\newcommand{\E}{\mathrm{E}}
\newcommand{\Eof}[1]{\E\left[#1\right]}

\newcommand{\prob}{\mathrm{P}}
\newcommand{\probOf}[1]{\prob(#1)}
\renewcommand{\vec}[1]{\mathbf{#1}}

\newcommand{\trans}{^{\mathsf T}}

\newcommand{\eqstop}{.}
\newcommand{\eqcomma}{,}
\newcommand{\defeq}{\coloneqq}
\newcommand{\disteq}{\, =_{\text{D}} \, }

\DeclareMathOperator*{\argmax}{arg\,max}

\newcommand{\ie}{\textit{i.e.}}
\newcommand{\eg}{\textit{e.g.}}

\newcommand{\quotes}[1]{``#1''}

\usepackage{todonotes}
\usepackage{ifthen}

\newboolean{draftversion}
\setboolean{draftversion}{false} 

\newboolean{longVersion}
\setboolean{longVersion}{false}  

\newcommand{\proofLocation}[2]{\ifthenelse{\boolean{longVersion}}{\text{#1}}{\text{#2}}}

\newcommand{\tdWasiur}[2][]{\ifthenelse{\boolean{draftversion}}{\todo[inline, color=blue!20, caption={2do}, #1]{\begin{minipage}{\textwidth-4pt}\emph{Remark Wasiur:}\\#2\end{minipage}}}{}}

\newcommand{\tdAmr}[2][]{\ifthenelse{\boolean{draftversion}}{\todo[inline, color=orange!20, caption={2do}, #1]{\begin{minipage}{\textwidth-4pt}\emph{Remark Amr:}\\#2\end{minipage}}}{}}

\newcommand{\tdSounak}[2][]{\ifthenelse{\boolean{draftversion}}{\todo[inline, color=red!20, caption={2do}, #1]{\begin{minipage}{\textwidth-4pt}\emph{Remark Sounak:}\\#2\end{minipage}}}{}}

\newcommand{\tdBastian}[2][]{\ifthenelse{\boolean{draftversion}}{\todo[inline, color=yellow!20, caption={2do}, #1]{\begin{minipage}{\textwidth-4pt}\emph{Remark Bastian:}\\#2\end{minipage}}}{}}

\newcommand{\tdGeneral}[2][]{\ifthenelse{\boolean{draftversion}}{\todo[inline, color=green!20, caption={2do}, #1]{\begin{minipage}{\textwidth-4pt}\emph{ToDO:}\\#2\end{minipage}}}{}}

\usepackage{url}

\usepackage{tabularx} 
\usepackage{xcolor}
\usepackage{xkeyval}
\input{tudcolours.def}

\usepackage{tikz}
\usetikzlibrary{arrows}
\usetikzlibrary{calc}

\usetikzlibrary{positioning,fit,shapes,backgrounds,circuits.logic.US}

\tikzstyle{server}=[circle, line width=0.5pt, rounded corners=0.1mm, draw=black!100, fill=tud3a!100]
\tikzstyle{vertex}=[circle, line width=1.5pt, draw=tud0d, fill=white]
\tikzstyle{dispatcher} =[and gate US, line width=0.5pt, draw=black!100, fill=tud0c!100]
\tikzstyle{dotbox} = [draw=white, fill=white, rectangle,  inner sep=10pt, inner ysep=20pt]

\tikzset{three_sided/.style={
		draw=none,rectangle, 
		append after command={
			[shorten <= -0.5\pgflinewidth]
			([shift={(-1.5\pgflinewidth,-0.5\pgflinewidth)}]\tikzlastnode.north west)
			edge([shift={( 0.5\pgflinewidth,-0.5\pgflinewidth)}]\tikzlastnode.north east)
			([shift={( 0.5\pgflinewidth,-0.5\pgflinewidth)}]\tikzlastnode.north east)
			edge([shift={( 0.5\pgflinewidth,+0.5\pgflinewidth)}]\tikzlastnode.south east)
			([shift={( 0.5\pgflinewidth,+0.5\pgflinewidth)}]\tikzlastnode.south east)
			edge([shift={(-1.0\pgflinewidth,+0.5\pgflinewidth)}]\tikzlastnode.south west)
		}
	}
}

%% file: sections/0_abstract.tex
Collaborative uploading describes a type of crowdsourcing scenario in networked environments where a device utilizes multiple paths over neighboring devices to upload content to a centralized processing entity such as a cloud service. Intermediate devices may aggregate and preprocess this data stream. Such scenarios arise in the composition and aggregation of information, \eg, from smart phones or sensors. We use a queuing theoretic description of the collaborative uploading scenario, capturing the ability to split  data into chunks that are then transmitted over multiple paths,
and finally merged at the destination. We analyze  replication and allocation strategies that control the mapping of data to paths and provide closed-form expressions that pinpoint the optimal strategy given a description of the paths' service distributions. Finally, we provide an online path-aware adaptation of the allocation strategy that uses statistical inference  to sequentially minimize the expected waiting time for the uploaded data. 
Numerical results show the effectiveness of the adaptive approach compared to the proportional allocation and a variant of the join-the-shortest-queue allocation, especially for bursty path conditions.

%% file: sections/1_introduction.tex

\section{Introduction}
\label{sec:intro}

Internet of Things (IoT) describes a world of 
heterogeneous devices, such as sensors and actuators that are connected through various communication technologies while carrying out everyday tasks. Crowdsourcing in the context of IoT often refers to interconnected devices that ubiquitously \emph{exchange} and \emph{aggregate } information to achieve complex goals. For example, live events can be  covered 
by composing many information streams originating from various mobile and fixed sources such as smart phones, audio/visual, and ambient sensors. Live events  include not only entertainment events but also emergency situations, such as security breaches  and attacks on civilians.

\begin{figure}
	\centering
	\includegraphics[width=\columnwidth]{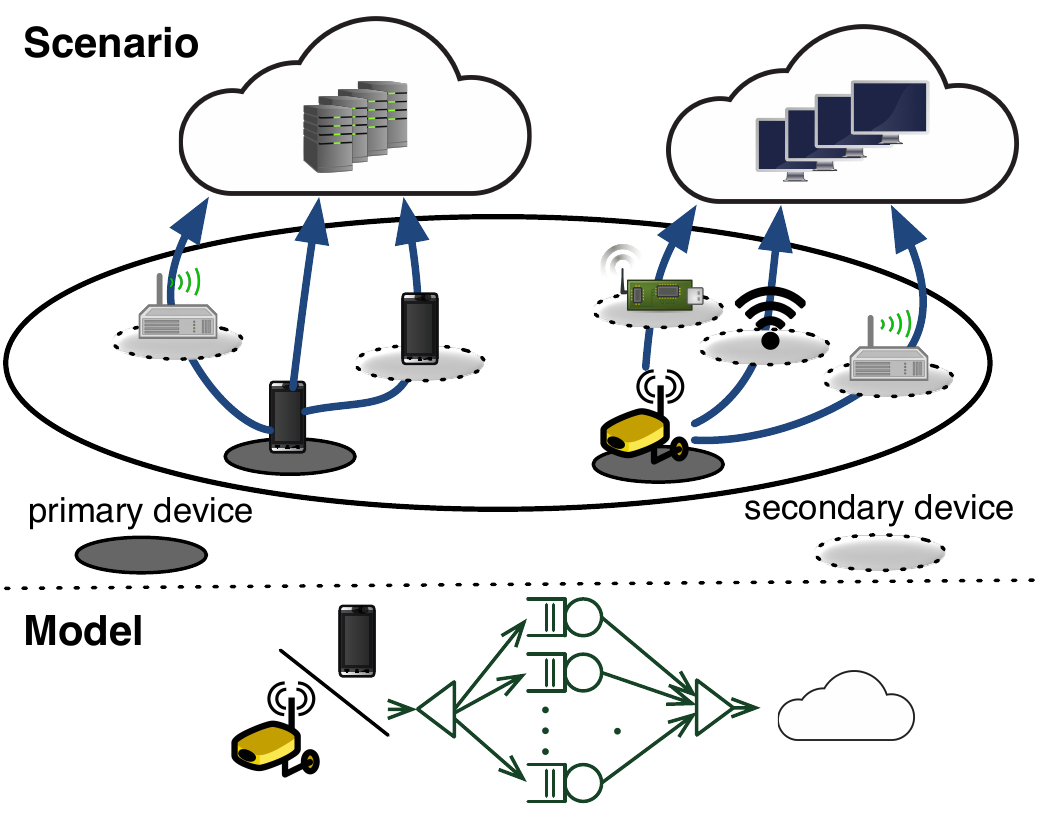}
	\caption{\label{fig:collaborative_uploading}%
	Collaborative uploading: A device uses neighboring devices and different paths to upload a data stream. The scenario is naturally captured by a Fork-Join queuing system.
	}
	\vspace{-5pt}
\end{figure}

A common feature of many of these crowdsourcing devices is the ability to simultaneously utilize different sets of wireless and wired communication technologies, such as WiFi, cellular, Ethernet and powerline communication, and to further recognize and simultaneously interact with surrounding devices. 
Fig.~\ref{fig:collaborative_uploading} shows different examples of collaborative uploading scenarios. As depicted, it is crucial to understand how a primary device can best utilize the parallel paths provided by secondary devices for uploading its data.


Modeling the collaborative uploading problem intrinsically includes scenario-specific challenges as shown by the heterogeneous examples in Fig.~\ref{fig:collaborative_uploading}. Nevertheless, we are interested in the uploading  performance, \eg, the time required to transfer a piece of data from a source device to a processing unit in an edge-cloud. An abstraction that enables powerful results on such performance measures is provided through queuing theory, as sketched in the bottom of Fig.~\ref{fig:collaborative_uploading}. This connection-layer abstraction enables the source to make intelligent decisions as to how to utilize the available and possibly heterogeneous paths by only considering their  latencies. 

Our goal in this work is to find optimal collaborative uploading strategies in crowdsourcing scenarios. 
We differentiate between intermittent (devices such as sensors sending data on a coarse time scale) and continuous collaborative uploading (devices continuously streaming video footage, \eg, using Facebook Live
). 
In optimizing performance metrics such as 
the uploading time, we also make a distinction between  the cases when devices possess full knowledge of the different path characteristics, 
and 
when they perform statistical inference. 


In this paper, we analyze replication and allocation strategies for collaborative uploading scenarios. We use a Fork-Join (FJ) queuing system (see Fig.~\ref{fig:collaborative_uploading}) that captures the ability to split data into chunks that are transmitted over multiple paths, and finally merged when all  chunks 
are received. 
Our contributions are summarized as: \textbf{1)} closed-form expressions for the mean upload latency in the intermittent uploading case, allowing a comparison between a replication and  an allocation (splitting) strategy. We  find optimal  strategies 
for given path latencies. 
In doing so, we also show numerical results suggesting near-optimality of the proportional allocation.
\textbf{2)} Online path-aware adaptation of the allocation strategy  based on statistical inference and stochastic gradient descent to sequentially minimize the expected waiting times in the continuous uploading case. We evaluate and compare the performance of our proposed adaptive strategy 
under various levels of path service burstiness.

The rest of this paper is organized as follows:
in Sec.~\ref{subsec:prelims}, we outline our modeling approach for the intermittent as well as the continuous uploading case. 
The intermittent case is then considered in detail in Sec.~\ref{subsec:single-file}. In Sec.~\ref{sec:streaming}, we pose the continuous uploading 
system as a queuing theoretic one and use stochastic gradient methods  to address the optimization of the allocation. In Sec.~\ref{sec:numerical_results}, we furnish an evaluation study of our proposed online algorithm. Finally, we discuss related work in Sec.~\ref{sec:related_work} and summarize our findings in Sec.~\ref{sec:discussion}.

%% file: sections/Model.tex
\section{Modeling approach}
\label{subsec:prelims}
Here, we present an overview of our 
 approach, which consists of \emph{(i)} defining an appropriate 
performance metric and \emph{(ii)} framing an appropriate optimization problem thereafter.

We characterize the intermittent case as one where the time intervals between two successive uploads are so large that there is \emph{no} self-induced queuing. Then, aspects such as cross-traffic can be described by means of the statistical properties of the path latencies alone.
A primary device uploading  data intermittently aims to minimize the upload latency, \ie, the time  until the data reaches the cloud.
Given multiple paths over secondary devices, the primary device may split the data into chunks that are transmitted or replicated over the available paths.
The upload latency being a stochastic quantity, it is natural to consider its mean as a performance metric and optimize it over all possible splitting/replication configurations.
In Sec.~\ref{subsec:single-file}, we express the  upload latency as 
an order statistic of the
individual upload times over the  different paths,  making the theory of order statistics a useful
tool in our analysis.

In the case of  continuous upload of a data stream, \eg,  a primary device uploading a live video to the cloud, there is a 
notion of waiting before each data chunk can be uploaded and hence, that of queuing.
We call the event of new data generation and passing by the application to the lower layers on the primary device, an arrival of a new data batch. Each data batch is split into  chunks of various sizes that are transported over several paths. Paths are characterized by a random service time required to transport the assigned chunks. Finally, the data batch reaches the cloud when all of its chunks are received.
Such systems are known as FJ queuing systems 
\cite{Rizk2015Sigmetrics,KhudaBukhsh2017infocom,baccelli1989fork}.

\section{Intermittent Collaborative Uploading}
\label{subsec:single-file}

In the following we consider the intermittent uploading case of data of size $K$ over $N$ possibly heterogeneous paths (\eg, sensor or monitoring devices uploading data on a coarse time scale).
Assume that the data can be divided into $N$ smaller chunks consisting of packets.
Then, every $\vec{k}=(k_1,k_2,\ldots,k_N) \in \diophantine{N}{K}  $  is a valid allocation vector, where  $k_i$ denotes the number of packets to be sent via path~ $i$ and   $\diophantine{N}{K} $ 
denotes the set of all non-negative integer solutions 
of the Diophantine equation $\sum_{i=1}^{N} k_i = K $, for $N, K\in \setOfNaturals$.  
We denote the random amount of time taken to transport the $j$-th packet out of the $k_i$ packets allocated to path $i$ by $D_{i,j}$. 
Here, $D_{i,j}$ may capture different phenomena that impact the transmission time over a path, such as resource allocation, transmission collisions, and retransmissions.
Assume that for each~$i \in \setN{N}$, with $\setN{N} \defeq \{1,2,\ldots, N \}$,  the random variables $D_{i,j}$'s are mutually independently distributed\footnote{Mutual independence, although not necessary for the subsequent analysis, is assumed for the sake of simplicity. In order to account for possible dependencies observed in real-world applications, one needs to additionally specify a correlation structure for these variables. This step is application-specific and is not easy in general. We do not attempt that in this paper. }.
Recall that the data consisting of $K$ packets can be reconstructed only after \emph{all} the packets have arrived. Therefore, the upload latency can be expressed as~$D \defeq \max ( D_1^{(k_1)}, D_2^{(k_2)}, \ldots, D_N^{(k_N)}  )$ where $D_i^{(k_i)} \defeq \sum_{j=1}^{k_i} D_{i,j}$ for $k_i>0$ denotes the amount of time taken by path~$i$ to transport $k_i$ packets, and by convention, $D_{i}^{(0)} \defeq  0 \, \forall i \in \setN{N}$.
The random variable~$D$ measures the total amount of time taken to transport \emph{all} the packets over $N$ different paths. 
In this work, we consider 
\begin{align*}
\psi(\vec{k}) \defeq  \Eof{ D}=  \Eof{ \max ( D_1^{(k_1)}, D_2^{(k_N)}, \ldots, D_N^{(k_N)}  )  } \, \eqcomma
\end{align*}
the expected upload time given an allocation $\vec{k}$, as our
performance metric. The density function of  $D_i^{(k_i)}$ is given by the $k_i$-fold self-convolution of the density function of $D_{i,j}$ due to independence.
Let us denote the  cumulative distribution function (CDF) of $D_i^{(k_i)}$ by $ F_i^{(k_i)} $. Stacking into a column vector $\vec{F}^{(\vec{k})} \defeq (F_1^{(k_1)},F_2^{(k_2)},\ldots, F_N^{(k_N)}  )\trans$, we  express the expected values of the order statistics of $D_1^{(k_1)},D_2^{(k_2)},\ldots, D_N^{(k_N)} $ as an operator~$\myMuoperator{}{}$ on $\vec{F}^{(\vec{k})}$ (see Remark~\ref{remark:moment_order_stat} in Appendix~\ref{subsec:order_stat}). Since $\psi(\vec{k})$ is 
the first moment of the $N$-th order statistic, 
we get 
\begin{align}
\psi(\vec{k})  = \myMuoperator{N}{  \vec{F}^{(\vec{k})}  } =  \sum_{j   \in \setN{N} }(-1)^{j +1 } \myDoperator{j}{ \vec{F}^{(\vec{k})} } \eqcomma
\label{eq:general_objective_func}
\end{align}
where  $\myMuoperator{N}{}$ and  $\myDoperator{j}{}$ are operators   defined in 
Appendix~\ref{subsec:order_stat}.  
The optimal allocation is found by minimizing $\psi$, \ie,
\begin{align}
\vec{k}_{\text{opt}} \defeq \argmin_{  \vec{k} \in \diophantine{N}{K} } \psi(\vec{k}) \eqstop
\label{eq:general_optimal_alloc}
\end{align}
Note that when the path characteristics are unknown, we can perform statistical inference\footnote{This is particularly important from an engineering perspective. The issue of statistical inference becomes more interesting  in the context of continuous upload. We show examples in  Sec.~\ref{sec:numerical_results}.}. In the following, we show some illustrative examples with computable $\vec{k}_{\text{opt}}$ before generalizing this allocation scheme to include replication strategies.

\begin{figure}
	\centering
	\begin{subfigure}[t]{0.49\columnwidth}
		\centering
		\includegraphics[width=\columnwidth]{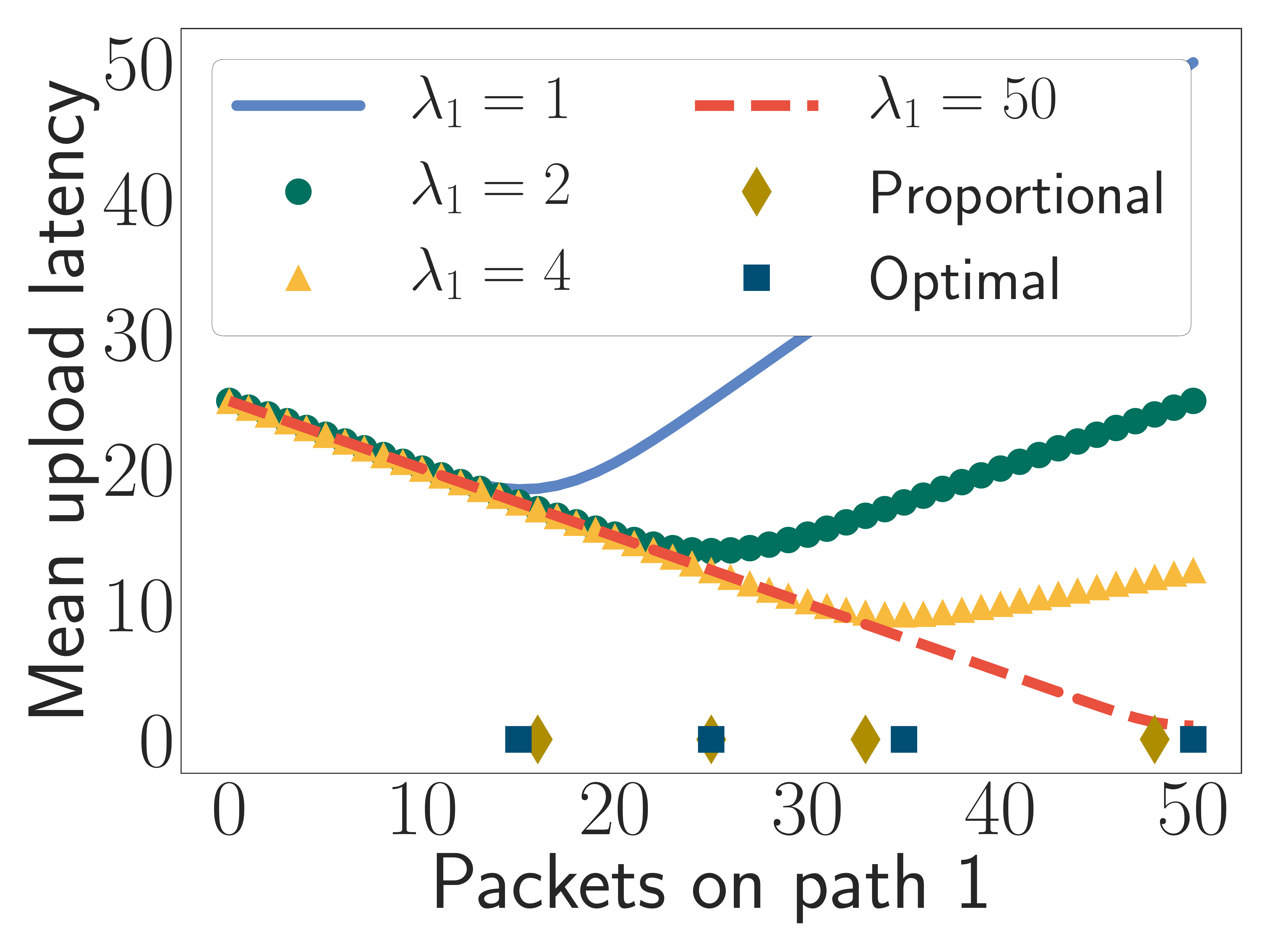}
	\end{subfigure}
	\begin{subfigure}[t]{0.49\columnwidth}
		\centering
		\includegraphics[width=\columnwidth]{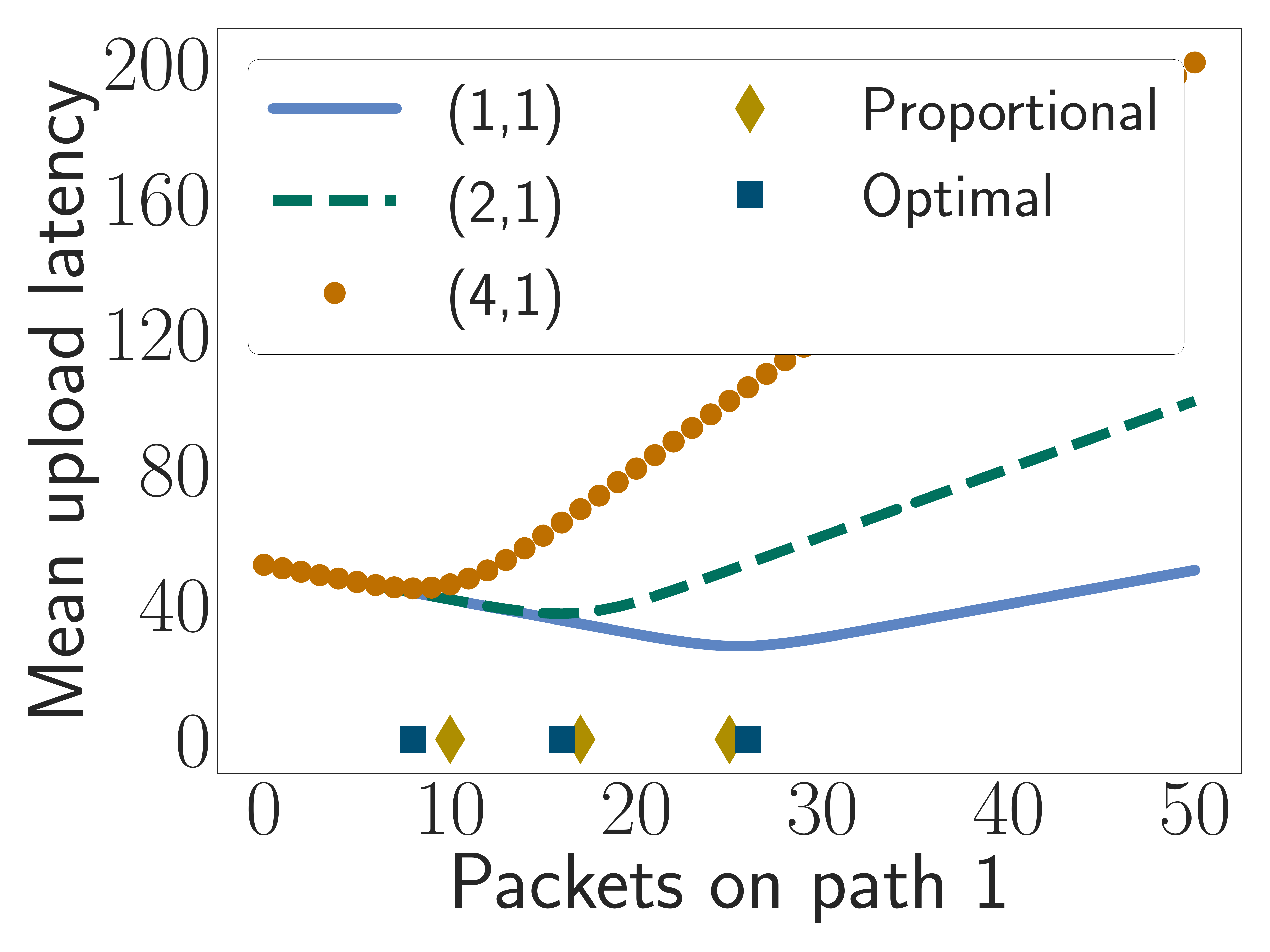}
	\end{subfigure}
	\caption{\label{fig:two-path}
	\textbf{Canonical two-path case:} We plot the mean upload latency as a function of the number of packets allocated to path~$1$ out of overall $50$ packets. \textbf{(Left)} Both paths have exponentially distributed delays.  The rate of the first path $\lambda_1$ is increased from 1 to 50, while that of the second path is fixed at~$\lambda_2=2$.
Note the shift in the optimal allocation as $\lambda_1$ increases.
		\textbf{(Right)} Path~$1$ has Weibull delay with (scale, shape) parameters given in the legend while path~$2$ has lognormal delay with parameters~$0$ and $0.25$.
Observe that the optimal allocation is indeed close to the proportional allocation.}
\end{figure}

\subsection{The canonical two-path case}
\label{subsec:two_path}
We consider the problem of finding the optimal allocation over two heterogeneous paths.
%
%
%
Let $\vec{k}  \in \diophantine{2}{K}$  denote our allocation. The corresponding upload latency  is given by
$D \defeq \max ( D_1^{(k_1)}, D_2^{(k_2)} )$   
and its mean is   $\psi(\vec{k}) = \myMuoperator{2}{\vec{F}^{(\vec{k})}} ={}    \Eof{  D_{1}^{(k_1)}   } +  \Eof{D_{2}^{(k_2)} }
- \int_{0}^{\infty} ( 1- F_1^{(k_1)} (x)   )( 1- F_2^{(k_2)} (x)  ) \, \differential{x}$. 
Suppose the packet latencies
$D_{1,j}$ and $D_{2,j}$ are exponentially distributed with rates $\lambda_1$ and $\lambda_2$.   
Then, setting $p=\frac{\lambda_1}{\lambda_1+\lambda_2}, q=1-p$, and $r=  \frac{1}{\lambda_1+\lambda_2}$, the expected upload time is
\begin{align*}
\psi(\vec{k}) =& \frac{k_1}{\lambda_1}+ \frac{k_2}{\lambda_2}
 - r \sum_{n_1=0}^{k_1-1}\sum_{n_2=0}^{k_2-1} \binom{n_1+n_2}{n_1} p^{n_1} q^{n_2} \eqstop
\end{align*}
	To minimize the above,  we derive the following relation through algebraic manipulation in \ifthenelse{\boolean{longVersion}}{Appendix~\ref{sec:AppendixB}}{\cite{KhudaBukhsh2017TechReport}}
	\begin{align*}
	\psi(k_1,k_2)  \gtreqless \psi(k_1+1,k_2-1)  \iff \frac{ I_{ p } (k_1, k_2  )     }{I_{ 1-p } (k_2-1, k_1+1  ) }  \gtreqless \frac{\lambda_2}{\lambda_1} \eqcomma
\end{align*}
	where $I_{x}(a,b)$ is the regularized $\beta$-function.
This allows finding the optimal allocation $\vec{k}_{\text{opt}}$
(see \ifthenelse{\boolean{longVersion}}{Appendix~\ref{sec:AppendixB}}{\cite[Appendix~B]{KhudaBukhsh2017TechReport}}). 
%
%

\ifthenelse{\boolean{longVersion}}{

\begin{figure}
	\centering
	\includegraphics[width=\columnwidth]{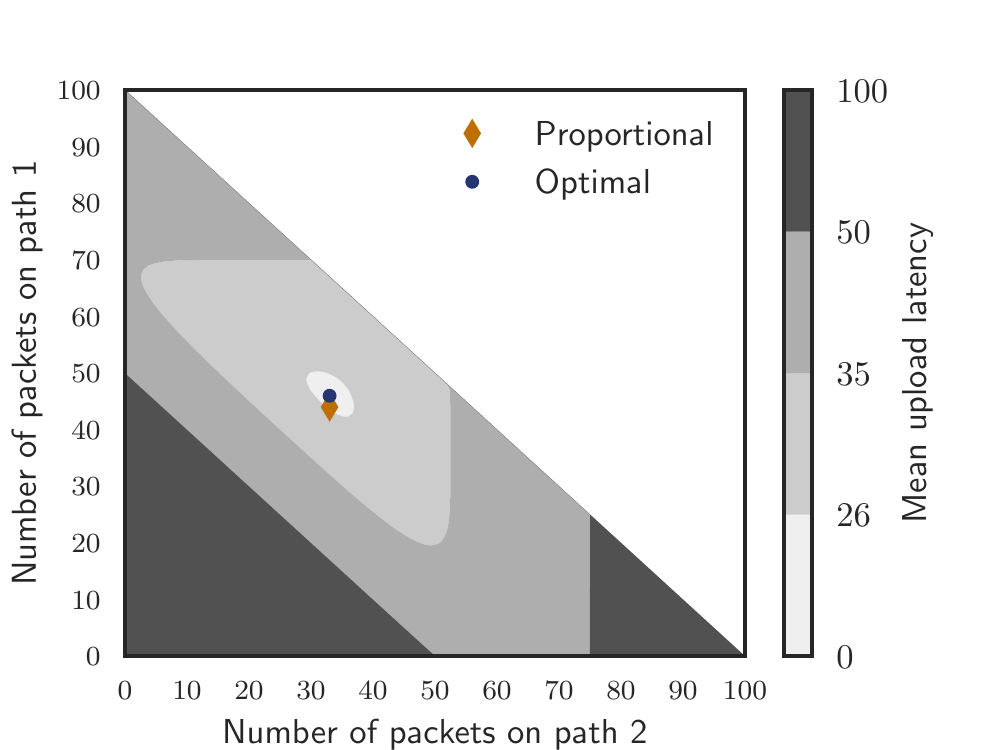}
	\caption{\label{fig:3path_exponential}%
	 Three heterogeneous paths with exponential delays with rates $2,1.5$ and $1$  are considered. The optimal allocation is centered at the innermost contour. The mean delay is high if the stronger paths (the first two) are grossly under-utilized (see bottom left corner). The proportional allocation is expectedly close to the optimal one. The number of packets considered in this example is $100$.}
\end{figure}
}{}

%
%
When 
$K$ is large, 
the optimal strategy can  be found by numerically solving the following nonlinear equation
\begin{align*}
\frac{  I_{ p  } (x, K-x ) }{  I_{ 1-p  } ( K-x-1, x+1 )  } - ( \frac{1}{p}  -1 ) =0 \eqstop
\end{align*}
In this case, the optimal allocation on path~$1$ is one of the two nearest integers producing a lower mean upload latency. 

In Fig.~\ref{fig:two-path}, we consider the canonical two-path scenario for different choices of path-specific delay distributions and show the mean upload latency as a function of the number of packets on path~$1$. 
For distributions not admitting a closed-form expression for the mean upload latency,
\eg, Weibull and lognormal, we  performed numerical integration.

\begin{figure}[t]
	\begin{subfigure}[t]{0.49\columnwidth}
		\centering
		\includegraphics[width=\columnwidth]{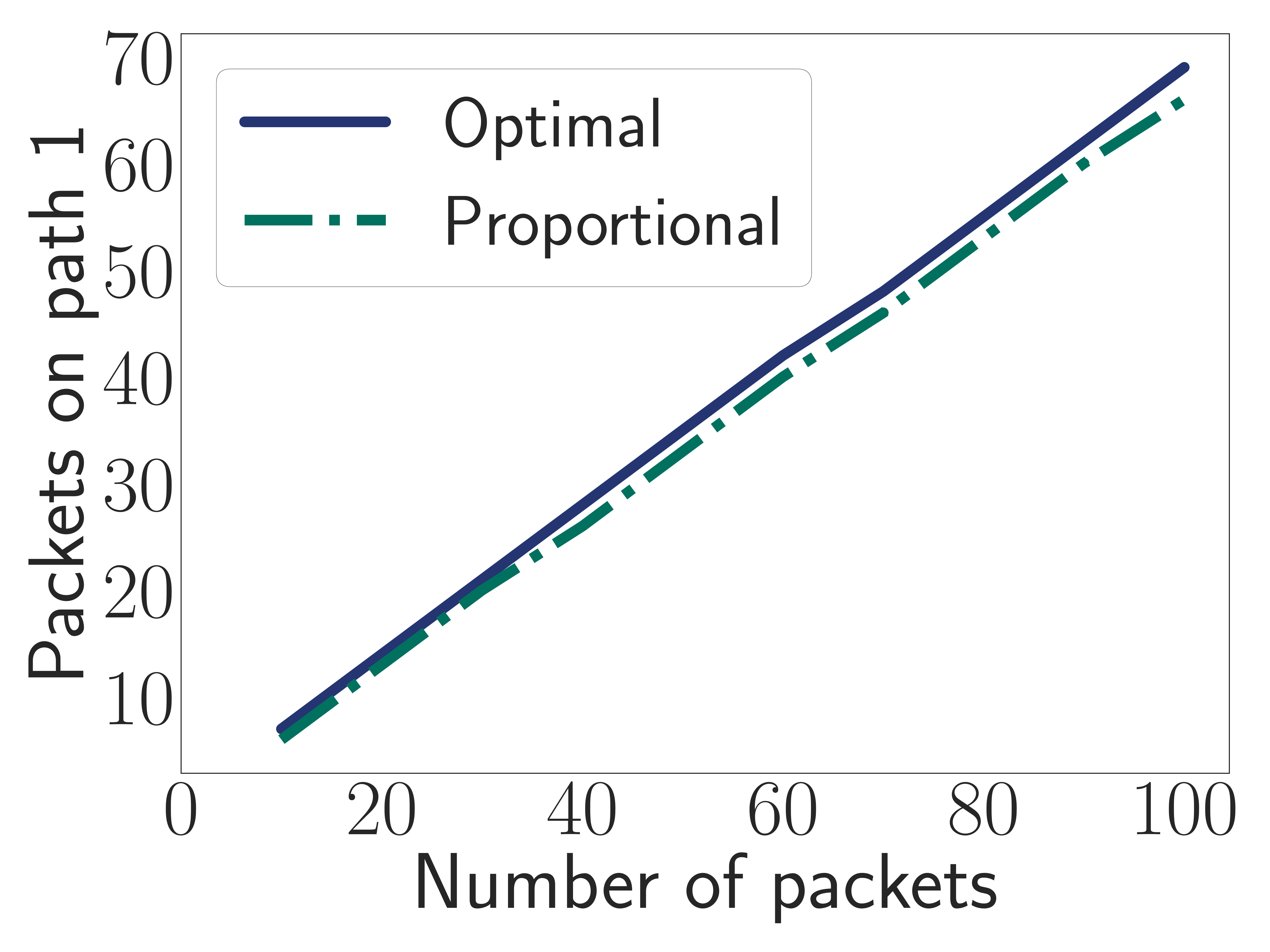}
	\end{subfigure}
	\begin{subfigure}[t]{0.49\columnwidth}
		\centering
		\includegraphics[width=\columnwidth]{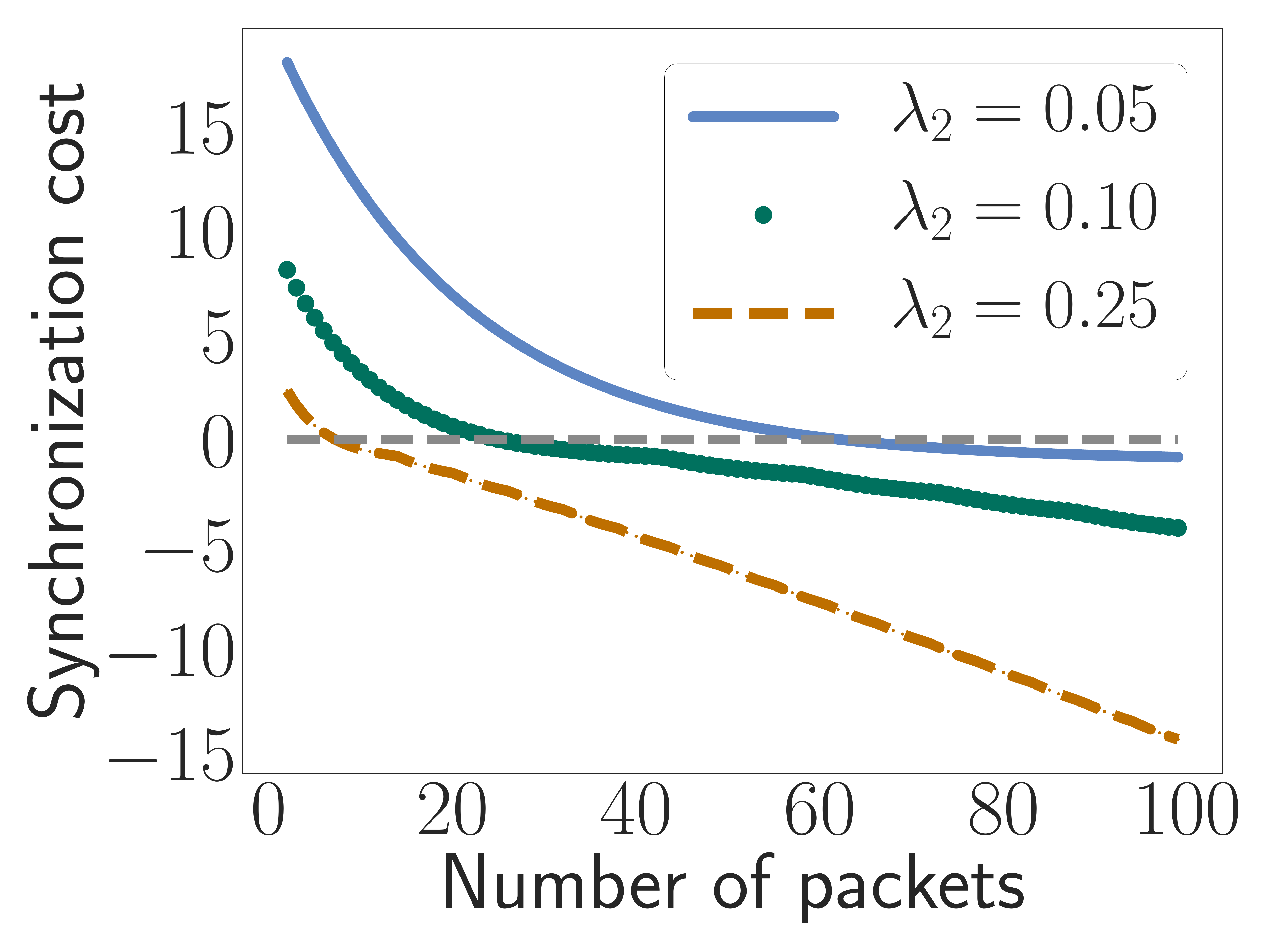}
	\end{subfigure}
	\caption{\label{fig:single_file_misc}%
		\textbf{(Left)} \textbf{Near-optimality of proportional allocation:} The number of packets allocated to path 1 vs. the overall number of packets (data size) for two exponentially distributed path latencies with parameters $\lambda_1=4$ and $\lambda_2=2$.
		\textbf{(Right)} \textbf{The synchronization cost as a function of data size:} We consider two  paths having exponential delays with rates $\lambda_1=1$ and $\lambda_2$ in the legend. Recall from \eqref{eq:syncho_cost} that positive (negative) synchronization cost implies superiority of replication (allocation).
 For large data sizes,  it is better to allocate than to replicate.
 However, if one of the paths is much slower compared to the other one, the synchronization cost is high and consequently, replication may become profitable.
		}
\end{figure}

\noindent\textbf{Near-optimality of proportional allocation: A comparison with \cite{Zhang2011Delay,Wen2007MinDelay}:} The two-path scenario has been studied in \cite{Zhang2011Delay,Wen2007MinDelay} for the exponential delay model. The authors, however, do not compute a closed-form expression for the mean upload latency and only provide the following upper bound, based on a Chernoff technique
\begin{align*}
\psi(k_1,k_2) \leq \max \left\{   \frac{k_1}{\lambda_1}, \frac{k_2}{\lambda_2}  \right\}   + \sqrt{2\pi } (  \sqrt{  \frac{k_1}{\lambda_1^2}  }  + \sqrt{  \frac{k_2}{\lambda_2^2}  }  ) \; \text{(due to \cite{Zhang2011Delay})} \eqstop
\end{align*}
Based on the above bound, the authors characterize the optimal allocation as being either the proportional allocation, \ie, $(k_1,k_2)= (Kp, K- Kp )$ or the winner-takes-it-all allocation, \ie, $(k_1,k_2) = (K,0)$. In contrast, we provide exact closed-form expression for the mean upload delay and find the optimal allocation $\vec{k}_{\textbf{opt}}$. 
Interestingly, we observe  near-optimality of the proportional allocation, \eg, as shown in Fig.~\ref{fig:single_file_misc} (left) for exponential path delays.
In Fig.~\ref{fig:two-path}, we see that similar conclusions hold for Weibull and lognormal delays as well.


%
%
%

\subsection{The N-path case with exponential delays}
\label{subsec:multipath_exp}
We next consider the  general case of $N$ paths available for uploading $K$ packets of data as sketched in Fig.~\ref{fig:collaborative_uploading}. 
Suppose the $i$-th path has  exponential delay with rate $\lambda_i$. The mean upload latency of the allocation $\vec{k}\in \diophantine{N}{K}$
is given by
\begin{align*}
\psi(\vec{k}) ={}&  \sum_{\substack{ S  \in  \{  A  \subseteq \setN{N} :  A \neq \phi   \} } }  (-1)^{\cardinality{S} +1}
 \sum_{ \substack{ 0 \leq n_i \leq {k_i-1}   :  i \in S } }   \left( \prod_{i \in S}  \frac{ \lambda_i^{n_i} }{n_i !} \right)  \\
& {} \times  \frac{  \Gamma( \sum_{i \in S} n_i +1  )   }{  (  \sum_{i \in S} \lambda_i  )^{\sum_{i \in S} n_i +1   } }    \, \eqstop
\end{align*}
The outer summation is carried out over all non-empty subsets of $\setN{N}$.
The derivation 
is provided in  \ifthenelse{\boolean{longVersion}}{Appendix~\ref{sec:AppendixB}}{\cite{KhudaBukhsh2017TechReport}}. The closed-form expression of $\psi(\vec{k})$ 
for more than two paths has not been provided before, to the best of our knowledge.




\ifthenelse{\boolean{longVersion}}{\begin{myExample}
	\label{example:exp3path}
	In particular, when there are three paths admitting exponential delays with parameters~$\lambda_1,\lambda_2$ and $\lambda_3$ respectively, the expression for mean delay corresponding to  an allocation~$\vec{k}=(k_1,k_2,k_3) \in \diophantine{3}{K}$ simplifies to
	\begin{align}
&	\psi(\vec{k}) \nonumber \\
=& \frac{k_1}{\lambda_1}+\frac{k_2}{\lambda_2}+\frac{k_3}{\lambda_3}   \nonumber  \\
	& - \frac{1}{\lambda_1+\lambda_2} \sum_{n_1=0}^{k_1-1}\sum_{n_2=0}^{k_2-1}   \frac{  (n_1+n_2)!  }{n_1! n_2!} \big(\frac{\lambda_1}{\lambda_1+\lambda_2}\big)^{n_1}\big(\frac{\lambda_2}{\lambda_1+\lambda_2}\big)^{n_2}  \nonumber \\
	& - \frac{1}{\lambda_2+\lambda_3} \sum_{n_2=0}^{k_2-1}\sum_{n_3=0}^{k_3-1} \frac{  (n_2+n_3)!  }{n_2! n_3!} \big(\frac{\lambda_2}{\lambda_2+\lambda_3}\big)^{n_2}\big(\frac{\lambda_3}{\lambda_2+\lambda_3}\big)^{n_3} \nonumber \\
	& - \frac{1}{\lambda_3+\lambda_1} \sum_{n_3=0}^{k_3-1}\sum_{n_1=0}^{k_1-1} \frac{  (n_3+n_1)!  }{n_3! n_1!} \big(\frac{\lambda_3}{\lambda_3+\lambda_1}\big)^{n_3} \big(\frac{\lambda_1}{\lambda_3+\lambda_1}\big)^{n_1}  \nonumber \\
	& + \frac{1}{\lambda_1+\lambda_2+\lambda_3} \sum_{n_1=0}^{k_1-1}\sum_{n_2=0}^{k_2-1} \sum_{n_3=0}^{k_3-1} \frac{  (n_1+n_2+n_3)!  }{n_1! n_2! n_3!} \big(\frac{\lambda_1}{\lambda_1+\lambda_2+\lambda_3}\big)^{n_1} \nonumber \\
	& \times   \big(\frac{\lambda_2}{\lambda_1+\lambda_2+\lambda_3}\big)^{n_2}  \big(\frac{\lambda_3}{\lambda_1+\lambda_2+\lambda_3}\big)^{n_3} \eqstop \label{eq:exp3path_alloc}
	\end{align}
	The exact expression of the mean delay above can be minimized to find the optimal allocation.
\end{myExample} }{}

\ifthenelse{\boolean{longVersion}}{ In Fig.~\ref{fig:3path_exponential}, we consider three heterogeneous paths with exponential delays with rates $\lambda_i$ equal to $\{2,1.5,1\}$ respectively.\ifthenelse{\boolean{longVersion}}{}{\footnote{In \cite{KhudaBukhsh2017TechReport}, we  provide a simplified expression for the mean delay.}}  The near-optimality  of the proportional allocation is observed here too (centered at the innermost contour).
}{
In  \cite{KhudaBukhsh2017TechReport}, we provide a simplified expression for the mean upload latency for three heterogeneous paths with exponential delays and optimize it.  The near-optimality  of the proportional allocation is observed for this example also.}


The allocation strategies discussed so far inherently impose a synchronization constraint at the destination. 
At a certain overhead, one  way to circumvent this synchronization constraint is replication, which we consider next.

\subsection{Replication strategies}
A basic replication strategy is to send the entire data over all available paths and take the first chunk that arrives at the destination. Replication strategies are known to reduce latency in some regimes~\cite{Vulimiri:2013:LLV}. However, an apparent drawback is their overuse of resources, \eg, higher energy consumption.
Roughly put, replication replaces the max operation (requiring the last chunk to arrive to complete the data at the receiver) with a min operation (taking the first to arrive at the receiver). However, the min operation is taken over elements that stochastically dominate the elements over which the max operation is taken. This poses an interesting trade-off: \emph{When should we  replicate, and not allocate?} 
%
%

In the basic replication case, the
upload latency is~$D \defeq \min ( D_1^{(K)}, D_2^{(K)}, \ldots, D_N^{(K)}  )$ where $D_i^{(K)} \defeq \sum_{j=1}^K D_{i,j}$, as before.
Our objective remains minimizing the mean upload latency 
\begin{align*}
\phi(N,K) \defeq \Eof{ D}  =  \Eof{ \min ( D_1^{(K)}, D_2^{(K)}, \ldots, D_N^{(K)}  )  }  = \myMuoperator{1}{\vec{F}^{(K \vec{\upsilon}  )}}  \, \eqcomma
\end{align*}
where $\vec{\upsilon}$ is an $N$-dimensional vector of all ones and $\vec{F}^{(K \vec{ \upsilon } )} = (F_1^{(K)}, F_2^{(K)}, \ldots, F_N^{(K)} ) $.
We  favor the replication strategy if $\phi(N,k)$ is smaller than the mean upload latency of \textbf{any allocation} $\vec{k} \in \diophantine{N}{K}$, \ie, if  $\phi(N,K)  \le   \min_{ \vec{k} \in \diophantine{N}{K}} \psi (\vec{k})$. 
In relation to the question of replication versus allocation, we introduce next the notion of synchronization cost.

\noindent \textbf{Synchronization cost:} 
Suppose all available paths are used for transmission and let
$\Lambda^*(N,K)\defeq  \{ \vec{k} = (k_1,k_2,\ldots, k_N) \in \diophantine{N}{K} \mid k_i >0 \quad \forall \quad  i \in \setN{N}  \}  $ denote the reduced set of valid allocations. Within $\Lambda^*$, an allocation  can be worse than a replication essentially because of the synchronization at the destination, \ie, because of some paths being much slower than others. To compare with a replication strategy, we define the synchronization cost given $N$ paths and data  size $K$ as
\begin{align}
\label{eq:syncho_cost}
\chi(N,K) \defeq & {}  \min_{\vec{k} \in \Lambda^*(N,K)} \psi (\vec{k}) - \phi (N,K)  \nonumber \\
=& {}   \min_{\vec{k} \in \Lambda^*(N,K)}   \myMuoperator{N}{\vec{F}^{(\vec{k})}}  -  \myMuoperator{1}{\vec{F}^{(K \vec{\upsilon}  )}} \eqstop
\end{align}
If $\chi$ is positive, replication yields smaller mean upload latency and hence, is preferred. If $\chi$ is negative, we prefer allocation over replication.
Intuitively, if the data size is  large, we expect the cost of redundancy  to be  high and $\chi$ to be negative. 


Consider the canonical two-path example with exponential delays from Sec.~\ref{subsec:two_path}. 
A straightforward computation of   $ \myMuoperator{1}{\vec{F}^{(K \vec{\upsilon}  )}}$ yields the following closed-form expression of the synchronization cost  defined in  \eqref{eq:syncho_cost},
\begin{align*}
\chi(2,K) = &{}  \min_{(k_1,k_2) \in \Lambda^*(2,K)}  \psi(k_1,k_2)
 - r \sum_{n_1=0}^{K-1}\sum_{n_2=0}^{K-1} \binom{n_1+n_2}{n_1} p^{n_1} q^{n_2} \eqstop
\end{align*}
In Fig.~\ref{fig:single_file_misc}, we show the synchronization cost as a function of the data size~$K$. As the data size increases the cost of redundancy worsens the performance of replication. Consequently, an allocation strategy 
is preferred for large data. However, the \emph{zero-crossing} data size seen in Fig.~\ref{fig:single_file_misc}, which marks the regimes where replication and allocation are more beneficial, shifts depending on path heterogeneity.

%


\subsection{Combined Allocation and Replication: An $(N,r)$-strategy}
\begin{figure}
	\centering
	\includegraphics[width=\columnwidth]{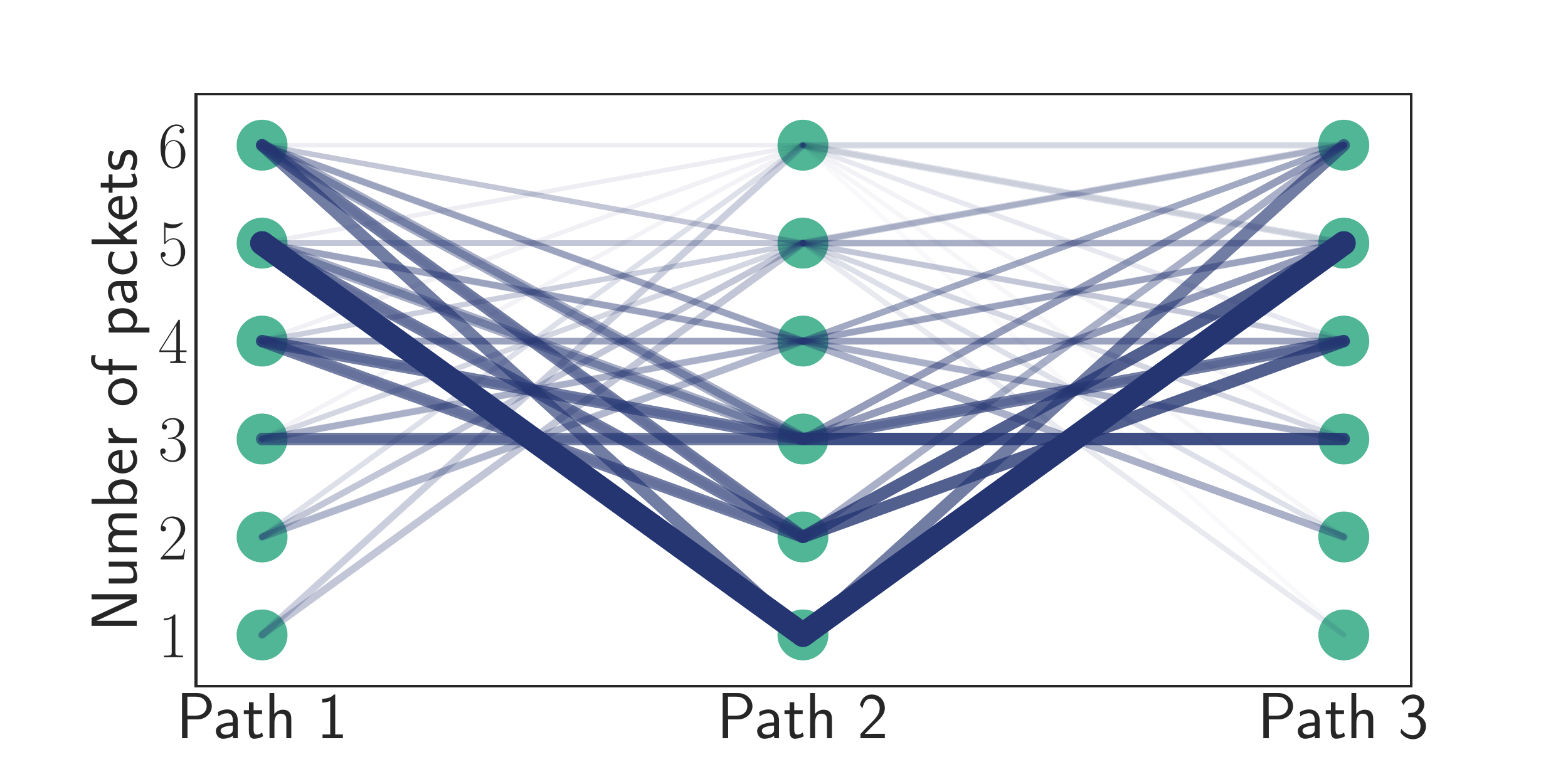}
	\caption{\label{fig:regret_gen_alloc}%
	\textbf{Optimal allocation by minimizing regret:} The lines 
	specify different $(3,2)$-allocations. For example, the line joining $1, 5,$ and $6$ corresponds to the allocation  $(1,5,6)$. 
Valid $(3,2)$-allocations require the combined size of any 
2 chunks to be 
at least the data size, here, 6.
The darkness/thickness of the shades is inversely proportional to the regrets defined in \eqref{eq:regret_defn}. The allocation $(5,1,5)$ (the thickest line), achieves zero regret and hence, is the optimal one. 
	We assume exponential delays with rates 
	$1,5$ and $10$ in order of increasing path indices. }
\end{figure}

Here, we  present  a variant of the  replication strategy,  called the $(N,r)$-strategy. An  $(N,r)$-strategy splits data of size $K$ into $N$ smaller chunks so 
that the data batch can be reconstructed from any $r$ out of the $N$ chunks. One of the ways to achieve such a splitting is to use Erasure codes, \eg, maximum distance separable (MDS) codes \cite{Joshi:2017:ERT}.  Note that an $(N,N)$-strategy corresponds to allocation and an $(N,1)$-strategy, to replication.
%
%
%
To formulate an $(N,r)$-strategy, we define
\begin{align*}
\genAllocation{N}{r}{K} \defeq  \{ \vec{k} \in \setN{K}^N \mid  \sum_{ i \in S } k_i \ge K \,  \forall \,  S \subseteqq [N] \, \eqcomma \cardinality{S} = r   \}  \eqstop
\end{align*}
We call a $\vec{k} \in \genAllocation{N}{r}{K} $ an $(N,r)$-allocation for data of size $K$. 
The data is received as soon as the first $r$ out of $N$ chunks arrive at the destination.  
Let the  order statistics corresponding to $D_1^{(k_1)},D_2^{(k_2)},\ldots, D_N^{(k_N)}$ be denoted by $C_1 \le C_2 \le \ldots \le C_N $. The mean upload latency for 
$\vec{k} \in \genAllocation{N}{r}{K}$ is $\eta_r (\vec{k}) \defeq  \Eof{C_r} =  \myMuoperator{r}{  \vec{F}^{(\vec{k})}   } $. 
In Appendix~\ref{sec:AppendixA}, we provide an example of $(N,r)$-allocations given three heterogeneous paths with  exponential delays.
For a fixed $r$, the optimal $(N,r)$-allocation is given by $\vec{k}_{\text{opt}}^{(r)} \defeq \argmin_{\vec{k} \in \genAllocation{N}{r}{K}  } \eta_r (\vec{k})  $. We can, however, further improve the performance by optimizing over~$r$.
To measure the performance of an allocation $\vec{k}$ compared to the optimal one, we define the regret of an $(N,r)$-allocation $\vec{k}$ as
\begin{align}
\gamma(\vec{k}) \defeq \eta_r (\vec{k})  -  \min_{r \in \setN{N}} \eta_r(\vec{k}_{\text{opt}}^{(r)} ) \eqstop
\label{eq:regret_defn}
\end{align}
In Fig.~\ref{fig:regret_gen_alloc}, we consider three heterogeneous paths with exponential delays. 
We find the optimal 
allocation by minimizing the regret. Interestingly, the optimal allocation is neither a $(3,1)$ replication, nor a $(3,3)$ allocation, but rather a $(3,2)$-allocation.



%% file: sections/Streaming_problem.tex
\section{Adaptive Collaborative Stream Uploading}
\label{sec:streaming}

\input{sections/Queuing_Interpretation}

\subsection{Adaptive Allocation}
\label{subsec:stoch_gradient}
We consider the problem of sequentially optimizing allocations for collaborative uploading of incoming data batches. The procedure is sketched in Fig.~\ref{fig:adaptive_allocation}.
Our adaptive allocation strategy seeks to minimize a sequence of cost functions
$(c_j)_{j\in \setOfNaturals}$
by  choosing a sequence of optimal allocation vectors
$(\vec{k}_j)_{j\in \setOfNaturals}$,
with $\vec{k}_j \defeq (k_{1,j},\dots,k_{N,j})$. An allocation vector $\vec{k}_j$ is a vector of integers, where the $n$-th entry corresponds to the number of packets (chunk size) transported over path $n$ and $\|\vec{k}_j\|_1=K_j$ is the size of the $j$-th data batch. Since optimization over integers is hard, we adopt the following standard relaxation: instead of an allocation vector, we optimize a proportion vector $\vec{x}_j \defeq (x_{1,j},\dots,x_{N,j})\in C$  corresponding to the $j$-th data batch, where $ C \defeq \{  \vec{y}  \in  [0,1]^N :  \|y \|_1 =1 \}$ is the set of valid proportions.  
The allocation vector $\vec{k}_j$ is found by 
taking the floor of first $N-1$ entries of $ \vec{x}_j K_j$  and subtracting their sum from $K_j$ 
to get the $N$-entry 
such that $\vec{k}_j  \in \diophantine{N}{K_j}$ (denote it as $\vec{k}_j \defeq \langle\vec{x}_j K_j\rangle_{\diophantine{N}{K_j}}  $). Adhering to our modeling approach in Sec.~\ref{subsec:prelims}, we choose the mean waiting time 
as our cost function.

Denote the service times of data batch $j$ on path $n$ and the inter-arrival times between data batch $j$ and $j +1$  by $S_{n,j}$ and $t_j$, respectively.  From a control theoretic perspective, we  treat $\vec{x}_{j-1}$ as our control variable to optimize the mean  of the following output 
$W_j$
\begin{equation}
\label{eq:fork_join_waiting_time}
W_j \defeq  \max \{ 0,   \underset{\begin{subarray}{c}
  n \in   \setN{N} \eqcomma
   k \in  \setN{j-1}
  \end{subarray}}{\max}  \{ \sum_{i  \in \setN{k} }  \left( x_{n,j-i} S_{n,j-i} -  t_{j-i} \right) \} \}  \eqstop
\end{equation}
The quantity in \eqref{eq:fork_join_waiting_time}   mimics the waiting time of $j$-th data batch in a Fork-Join system \cite{KhudaBukhsh2017infocom,Rizk2015Sigmetrics} with a diminishing rounding error for increasing batch sizes. 
The $j$-th cost function is 
\begin{align}
	c_j(\vec{x}_j)   \defeq \Eof{W_{j+1}} \eqcomma \label{eq:cost_func_defn}
	\end{align}
 We minimize the cost functions sequentially in an online fashion using gradient descent methods as they achieve a bounded regret in an online convex programming scenario~\cite{Zinkevich2003}.

As data batches are passed from the application on the primary device to be split over multiple secondary devices 
(Fig.~\ref{fig:collaborative_uploading}), we assume the $j$-th inter-arrival time $t_j$  is  known to the scheduler before employing the next proportion $\vec{x}_{j+1}$, given $ \{  \vec{x}_i \}_{ i \le j  } $.  
Using Monte Carlo (MC) methods we can calculate an unbiased estimate of the cost function for each data batch $j$. Since \eqref{eq:fork_join_waiting_time} is a piecewise linear function, which is non-smooth, we use an unbiased estimate of a subgradient $\vec{\hat{g}}_j$ of the $j$-th cost function $c_j$ in \eqref{eq:cost_func_defn}, to perform gradient descent. The definition of a subgradient is given in \ifthenelse{\boolean{longVersion}}{Appendix~\ref{sec:AppendixC}}{\cite{KhudaBukhsh2017TechReport}} and 
\cite{Poljak1987}.

\paragraph*{Gradient descent for data allocation} As shown in Fig.~\ref{fig:adaptive_allocation}, we 
update 
the
proportion $\vec{x}_{j}$ using gradient descent by
 \begin{equation}
 \vec{x}_{j+1}= \projection{ C }{\vec{x}_j+\eta \vec{\hat{g}}_j(\vec{x}_j)}
 \label{eq:greedy_projection}
 \end{equation}
where $\vec{\hat{g}}_j(\vec{x}_j)$ is an unbiased estimate of the subgradient of 
$c_j$ in \eqref{eq:cost_func_defn} evaluated at the current $\vec{x}_j$. Here, $\eta$ is the static learning rate  controlling the step size of the subgradient,  and $\projection{ C  }{.}$ denotes the Euclidean projection operator \cite{Wang2013}, projecting the gradient update onto the set of feasible proportions $C$. 

The update equation
\eqref{eq:greedy_projection}
ensures bounded regret with an unbiased estimate of the subgradient $\vec{\hat{g}}_j$ \cite{Flaxman2005}, which we obtain using Monte Carlo  methods. 
%
We resample service times $M$ times to estimate the subgradient.
The $m$-th sample for the  service times up to data batch $j$ at each path is denoted by $\{s_{n,i}^{(m)}\}_{n \in \setN{N}, i \le j}$. 
Then, the MC estimate of the subgradient is  
\begin{align}
  \hat{\vec{g}}_j(\vec{x}_j) = {} &  \frac{1}{M}\sum_{m \in \setN{M}} \Big( s_{n,j}^{(m)} \vec{e}_n   \indicator{   s_{n_m^*,k_m^*}^{(m)} \vec{e}_{n_m^*}\trans   \vec{x}_j \nonumber \\
&{}  +  \sum_{i=1}^{k_m^*-1} x_{n_m^*,j-i}s_{n_m^*  ,j-i}^{(m)}- \sum_{i=0}^{k_m^*-1}  t_{j-i}   >0   }   \Big) \eqcomma \label{eq:subgradient_sample}
\end{align}
%
where  $\vec{e}_n$ is the $n$-th unit vector, and $\indicator{.}$ is the indicator function.
The maximizers $n_m^*$ and $k_m^*$ 
can be found by 
\begin{align}
  (n_m^*,k_m^*)=  \argmax_{ \substack{  n \in \setN{N},\\ k \in \setN{j}   }  } \left\{  s_{n,j}^{(m)} \vec{e}_n\trans+ \sum_{i=1}^{k-1} x_{n,j-i}s_{n,j-i}^{(m)}-\sum_{i=0}^{k-1} t_{j-i} \right\} \eqstop
  \label{eq:subgradient_maximizer}
\end{align}
\ifthenelse{\boolean{longVersion}}{
 Note that due to ergodicity, queuing systems such as the one described in \eqref{eq:fork_join_waiting_time} possess regeneration points at the beginning of every busy period, i.e., at the last time point when the queue was empty. This reduces the history, i.e., the number of samples, required for calculating the subgradient substantially. 
 }{}
Algorithm \ref{alg:gd} describes the adaptive allocation method. In the next section, we describe the inference procedure required for our adaptive allocation.

\begin{algorithm}[H]
   \caption{Adaptive Allocation}
 \begin{algorithmic}[1]
   \State $j=1$, $\vec{x}_1 \gets \frac{1}{N}$ and $\vec{k}_1 \gets \langle\vec{x}_1 K_1\rangle_{\diophantine{N}{K_1}}$ //Initialization
   \State Draw M samples of $S_{n,j}$  at each batch arrival
   \ForAll{$m\in \setN{M} $}
  \State Calculate $  (n_m^*,k_m^*)$ using \eqref{eq:subgradient_maximizer}
   \EndFor
 \State Estimate the subgradient $\hat{\vec{g}}_j (\vec{x}_j)$ using \eqref{eq:subgradient_sample}
   \State $\vec{x}_{j+1} \gets \projection{C}{  \vec{x}_j+\eta \hat{\vec{g}}_j (\vec{x}_j)  }$, $\vec{k}_{j+1} \gets \langle\vec{x}_{j+1}K_{j+1}\rangle_{\diophantine{N}{K_{j+1}}}$  
   \State $j \gets j+1$

 \end{algorithmic} \label{alg:gd}
\end{algorithm}

\input{sections/Inference_part}

%% file: sections/Queuing_Interpretation.tex

Now, we analyze collaborative uploading for continuous data streams using an FJ queuing model. An example scenario is the continous upload of video data using multiple paths, as depicted in Fig.~\ref{fig:collaborative_uploading}. We first consider a rigid allocation strategy based on known probabilistic bounds on the steady-state waiting times before proposing an adaptive allocation scheme  
based on stochastic gradient descent.


\subsection{Rigid allocation  based on steady-state bounds}
Following \cite{Rizk2015Sigmetrics}, we define the waiting time of an incoming data batch as the amount of time it  waits until the last of its chunks starts getting uploaded.  Consider the steady-state waiting time $W$ (precise definition given in \cite{Rizk2015Sigmetrics} and for the sake of completeness, also in \ifthenelse{\boolean{longVersion}}{Appendix~\ref{sec:AppendixA}}{\cite{KhudaBukhsh2017TechReport}}). It is hard to find out the 
distribution of the steady-state waiting times in closed form (see \cite{Rizk2015Sigmetrics,Baccelli89}). One approach is to compute tight upper bounds on the tail probabilities.
Following \cite{KhudaBukhsh2017infocom}, for  a given  allocation~$\vec{k} \in \diophantine{N}{K}$ and independent service times, we get
\begin{align}
\label{eq:bound_heterogeneous}
 \probOf{ W \geq \sigma }  \leq {}& \myExp{ - \tilde{\theta}  \sigma  }  \sum_{i \in \setN{N} } \myExp{ -  ( \theta_i -  \tilde{\theta}   )\sigma }  \eqcomma
\end{align}
where $\theta_i>0$ is
given by a condition
involving the Laplace transforms of the inter-arrival times and the service times for $k_i$ packets and $\tilde{\theta} \defeq \min_{i \in \setN{N} } \theta_i$. Here, $\tilde{\theta}$ is the effective decay rate of the tail probability in the sense of large deviations principle, and  
assesses the quality of a given allocation (the higher the decay rate, the better).

Reducing the waiting times is equivalent to maximizing the effective decay rate. Treating $\tilde{\theta}$ as a function of the allocation,
the optimal allocation is given by
\begin{align}
\vec{k}_{\text{opt}} \defeq \argmin_{\vec{k} \in \diophantine{N}{K} } \tilde{\theta} (\vec{k}) \eqstop
\label{eq:general_optimal_alloc_bound}
\end{align}

\begin{figure}
	\centering
	\includegraphics[width=\columnwidth]{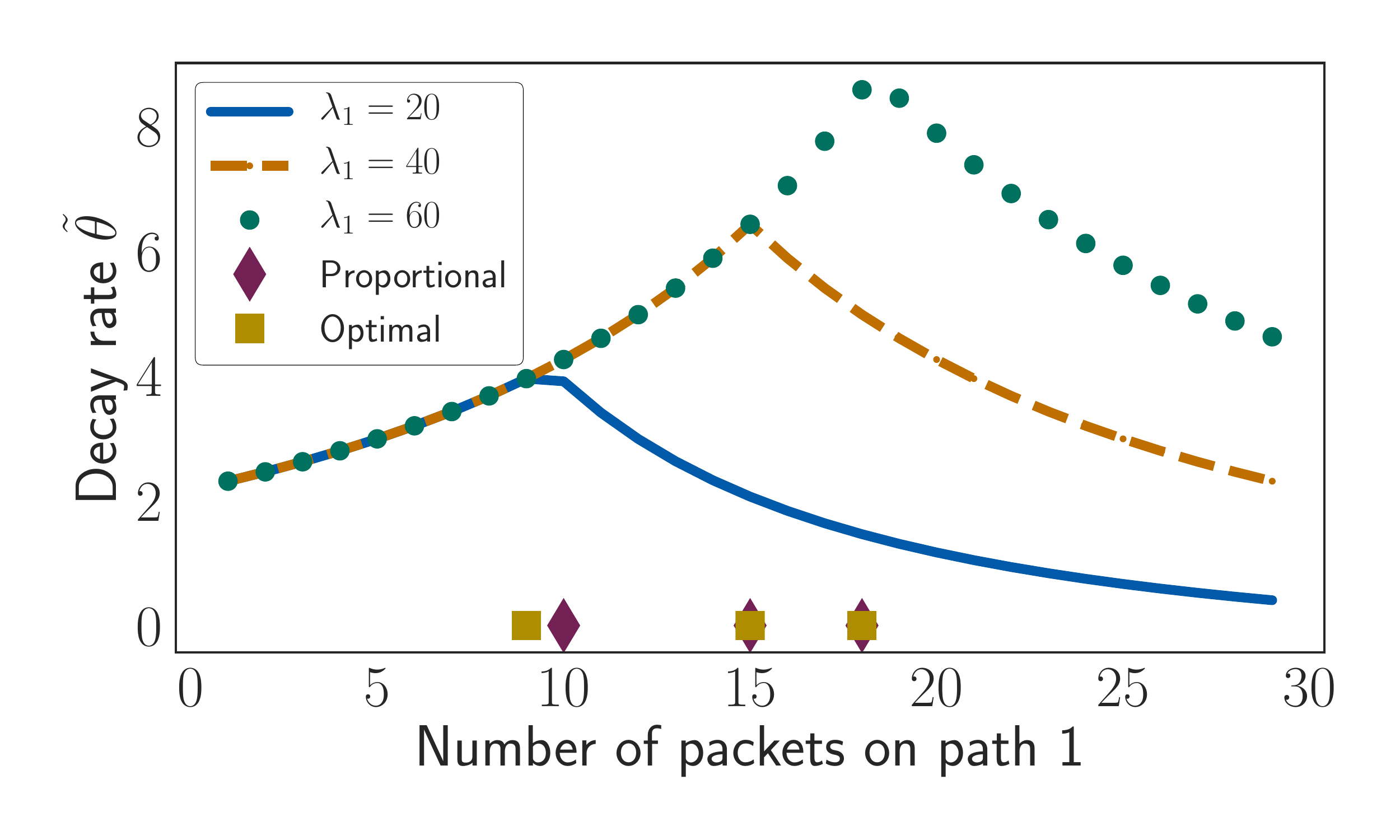}
	\caption{	\label{fig:theta_opt}%
  \textbf{Canonical two-path scenario for collaborative stream uploading:} The effective decay rate $\tilde{\theta}$ from \eqref{eq:bound_heterogeneous} as a function of the number of packets sent via path~$1$ out of overall 30 packets. Both paths have exponential delays. We vary the rate of the first path $\lambda_1$ as shown in the legend and keep the rate of the second path fixed at $\lambda_2=40$. The inter-arrival times are exponentially distributed with rate $0.5$. Here too, we observe  near-optimality of the proportional allocation.
  }
     \vspace{-5pt}
\end{figure}

	In Fig.~\ref{fig:theta_opt}, we revisit the canonical two-path scenario with exponential delays\ifthenelse{\boolean{longVersion}}{ (derivation in  Example~\ref{example:decay_rate_exp} in  Appendix~\ref{sec:AppendixA})}{ (derivation in \cite[Example 3]{KhudaBukhsh2017TechReport})}.  
   Plotting the effective decay rate $ \tilde{\theta}$ as a function of the number of packets on path~$1$,
    we find  the optimal allocation (yielding the largest decay rate). We also observe the near-optimality of the proportional allocation. 


	The approach in \eqref{eq:general_optimal_alloc_bound} is  convenient 
	because of its  simplicity. However, it has a number of drawbacks.
 Apart from the exponentially growing search space for the optimal allocation, the approach 
   is valid for the steady-state waiting time only. In many applications, the transient behavior is important. 
   The approach in \eqref{eq:general_optimal_alloc_bound} does not allow for adaptation as it ignores the current state of the system (the number of chunks already on each path). In a realistic setup with changing environment (\eg, Markov-modulated paths' services), the ability to adapt is crucial. Keeping this in mind, we propose an adaptive allocation scheme in the next section.

\ifthenelse{\boolean{longVersion}}{
\begin{figure}
  \includegraphics[width=\columnwidth]{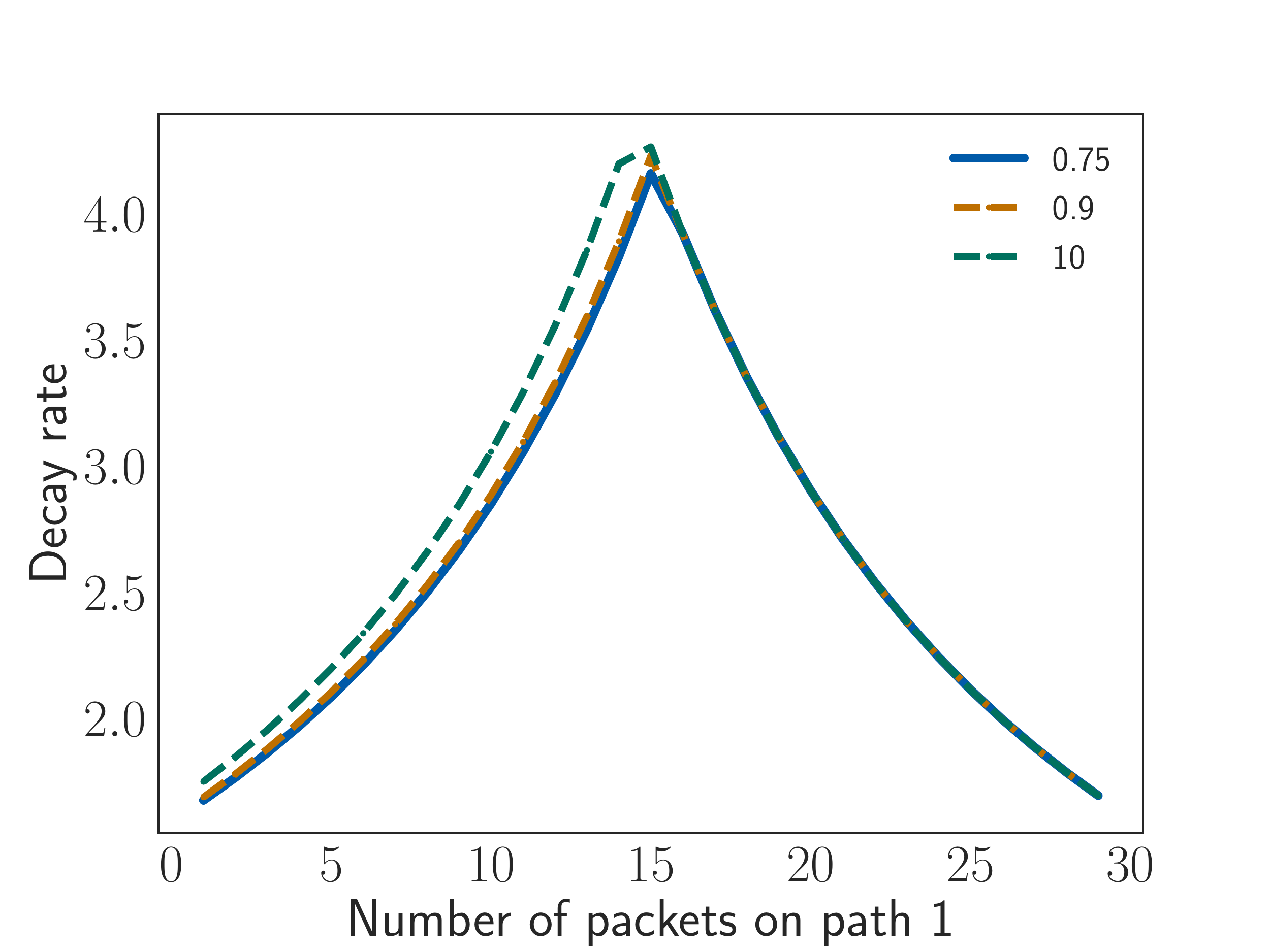}
    \caption{	\label{fig:theta_opt_gamma}%
    The decay rate achieved as a function of the number of packets sent via path~$1$ in the canonical two-path scenario. Both the paths are assumed to have gamma delays with the same mean but different variance (achieved by scaling up the parameters of the gamma distribution). The gamma distribution has parameters $40,2$ and the scale-up parameters are shown in the legend. The data size in this case is $30$. The arrival rate is assumed to be $0.15$. The decay rate is maximized at the corresponding proportional allocation. We also observe that smaller variance gives higher decay rate.}
\end{figure}
}{}


%% file: sections/Inference_part.tex
\subsection{Inference for service time processes}
\label{sec:Inference_service_time}

\begin{figure}
	\centering
	\includegraphics[width=\columnwidth]{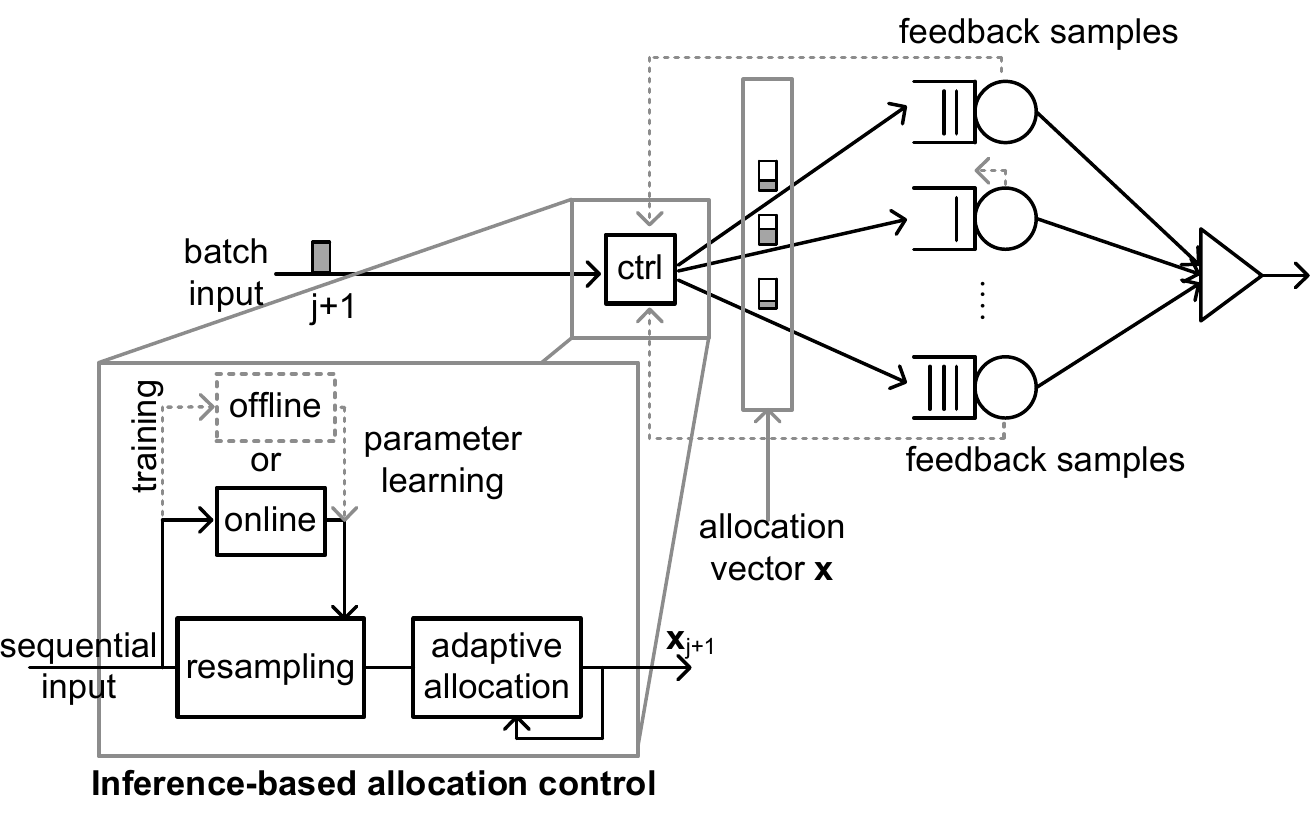}
	\caption{\label{fig:adaptive_allocation}%
	At the arrival of each data batch, the controller first computes an MC estimate of the subgradient using  \eqref{eq:subgradient_sample} before finding the next allocation vector using
\eqref{eq:greedy_projection}. In order to get the MC estimate, we resample the service times $M$ times. If unknown, we infer the latent state of the Markov chains that describe the paths' service in an online fashion, while other model parameters are learned offline using training samples. Observed service times serve as feedback to the controller.}
\end{figure}

Since the adaptation of allocations requires samples of the service times (Step 2 of Algorithm~\ref{alg:gd}),
we provide here 
illustrative resampling schemes for independent and identically distributed (i.i.d.) and Markov-modulated service times.
\paragraph*{Exponential i.i.d. service times} Assume i.i.d service times at the different paths. 
Suppose the service times at one of the paths are exponentially distributed with rate parameter~$\lambda$  so  that the distribution of $S_{n,j}$ is gamma with shape parameter $K_j$, and rate $ \lambda$.  The predictive distribution of a new sample of service time $ S^*_{n,j} $ 
based on a set of observed data   $\{s^{\prime}_{n,i} \}_{i \le j}$
is found by integrating out the parameter $\lambda$, \ie,  by  $ \int   p( S^*_{n,j}  | K_j, \lambda) p(\lambda |  \{s^{\prime}_{n,i} \}_{i \le j} ) \mathrm{d}\lambda$. If we  assume a conjugate prior $p(\lambda)$ on $\lambda$, namely, a gamma distribution with (hyper)-parameters $K_0$ and $\lambda_0$, the posterior  $p(\lambda | \{s^{\prime}_{n,i} \}_{i \le j} )$ is a gamma distribution
with (posterior) parameters $K_{\mathrm{p}}=K_0+\sum_{i=1}^j K_{n,i}$ and $\lambda_{\mathrm{p}}=\lambda_0+\sum_{i=1}^j s^{\prime}_{n,i}$. The derivation steps 
are provided in \ifthenelse{\boolean{longVersion}}{Appendix~\ref{sec:AppendixC}}{\cite{KhudaBukhsh2017TechReport}}. In order to sample 
from the predictive distribution, 
we first sample 
$\lambda$ from the posterior distribution 
and then 
sample from $p(S^*_{n,i}|K_i, \lambda)$ conditioned on $\lambda$ (step 2 in Algorithm~\ref{alg:gd}). 
Repeating this procedure iteratively, we get the following  update equations for the
posterior parameters 
$K_{\mathrm{p}}^{(i)}=K_{\mathrm{p}}^{(i-1)}+K_{n,i} \eqcomma \, \lambda_{\mathrm{p}}^{(i)}=\lambda_{\mathrm{p}}^{(j-1)}+s^{\prime}_{n,i}$,
with $K_{\mathrm{p}}^{(0)}=K_0$ and $\lambda_{\mathrm{p}}^{(0)}=\lambda_0$.


\ifthenelse{\boolean{longVersion}}{
\begin{figure}
	\centering
	\begin{tikzpicture} [inner sep=0cm, minimum size = 1.25cm]
		\draw [line width=1pt, draw=tud0d, fill=tud0b!40] (-2.5,-1) rectangle  +(8.5,2);
		\node[below] at (4.7,1.2){Hidden layer};

		\draw [line width=1pt, draw=tud0d, fill=tud0b!40] (-2.5,-3.75) rectangle  +(8.5,2);
		\node[below] at (4.6,-1.5){Observed layer};

			\node(vertex1) [vertex] {$Z_t$} ;
			\node(vertex2) [vertex, right=1.5cm of vertex1] {$Z_{t+1}$} ;
			\node(vertex3) [vertex, below=1.5cm  of vertex1] {$X_t$} ;
			\node(vertex4) [vertex, below =1.5cm of vertex2] {$X_{t+1}$} ;
			\coordinate[right=1.5cm  of vertex2] (d1);
			\coordinate[left = 1.5cm of vertex1] (d2);

			\draw [-latex, line width=1pt,in=180, out=0] (vertex1) to node [minimum size=0.3cm, xshift=0.1cm, yshift=0.25cm] {} (vertex2);
			\draw [-latex, line width=1pt,in=90, out=-90] (vertex2) to node [minimum size=0.3cm, xshift=0.1cm, yshift=0.25cm] {} (vertex4);
			\draw [-latex, line width=1pt,in=90, out=-90] (vertex1) to node [minimum size=0.3cm, xshift=0.1cm, yshift=0.25cm] {} (vertex3);
			\draw [-latex, line width=1pt] (d2) to node [minimum size=0.3cm, xshift=0.1cm, yshift=0.25cm] {} (vertex1);
			\draw [-latex, line width=1pt] (vertex2) to node [minimum size=0.3cm, xshift=0.1cm, yshift=0.25cm] {} (d1);
			\end{tikzpicture}
			\caption{\label{fig:Markov_modulated}
			A probabilistic graphical model representation of a Markov modulated process. The modulating Markov chain is denoted by $Z_t$, and the modulated process, by $X_t$. For our purpose, the modulated processes are the inter-arrival times for the arrival process and the service times. To be specific, we assume the modulating Markov chain modulates the parameters of inter-arrival and service time distributions.
			}
\end{figure}
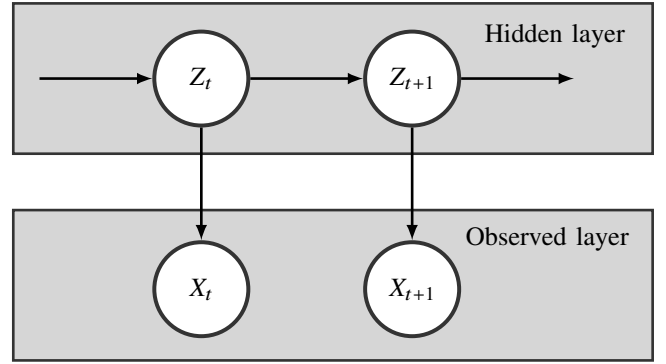
}{}

\paragraph*{Markov modulated service times}  In this case, we assume that the service times are instances of a Markov modulated exponentially distributed variable. 
For a sequence of service times, we first find maximum-likelihood estimates (MLE) of the underlying parameters, \eg, the initial distribution, transition matrix and the rates. Since 
the MLE computation is  expensive, 
this estimation process is executed offline. The derivation of the MLE 
is provided in \ifthenelse{\boolean{longVersion}}{Appendix~\ref{sec:AppendixC}}{\cite{KhudaBukhsh2017TechReport}}.
Next, we condition on these parameters to calculate the \emph{maximum a posteriori} (MAP) estimate   
of the current hidden state of the Markov chain. 
We do this online using the Viterbi algorithm \cite{Ephraim2002HMM}. Having inferred the hidden state, we resample the service times (step 2 in Algorithm~\ref{alg:gd})
online by conditioning on the hidden state and the parameters estimated in the first step.

%% file: sections/Numerical.tex
\section{Numerical Evaluation}
\label{sec:numerical_results}

	\begin{figure*}[t!]
	\centering
		\begin{subfigure}[t]{0.32\textwidth}
			\centering
			\includegraphics[width=\textwidth]{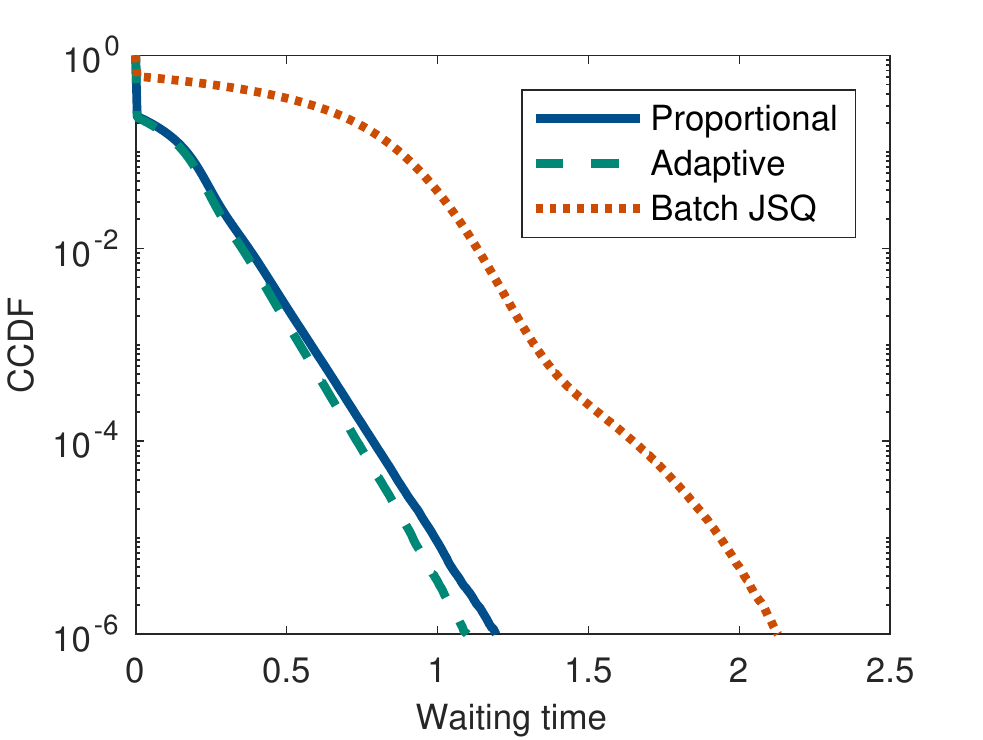}
		\end{subfigure}
		\hfill
		\begin{subfigure}[t]{0.32\textwidth}
			\centering
			\includegraphics[width=\textwidth]{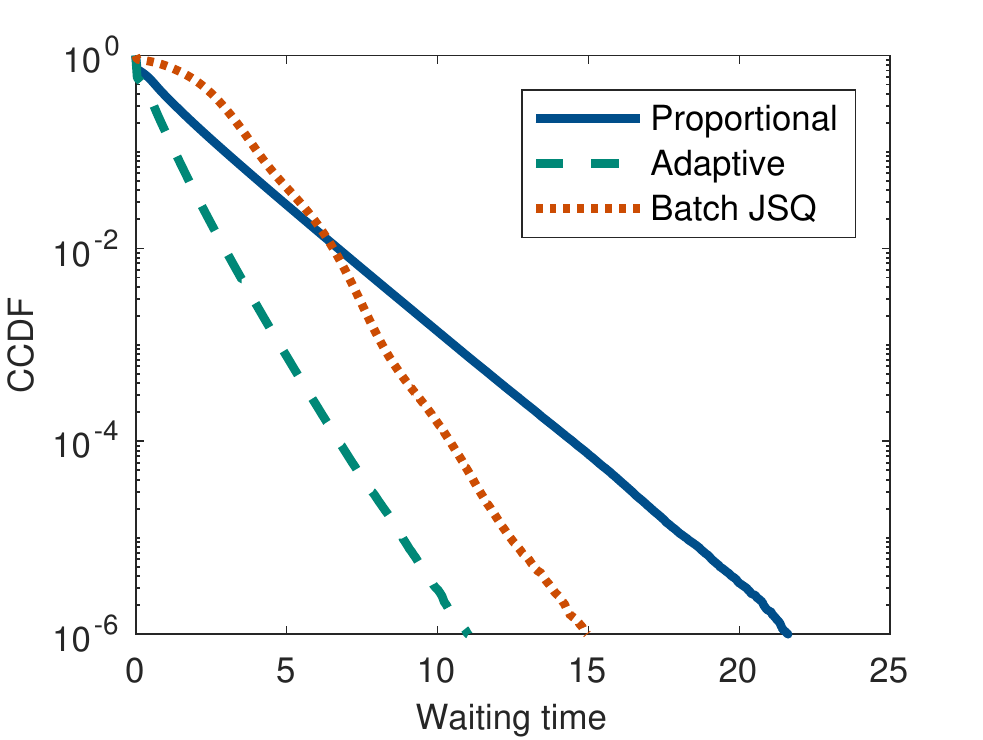}
		\end{subfigure}
		\hfill
		\begin{subfigure}[t]{0.32\textwidth}
			\centering
			\includegraphics[width=\textwidth]{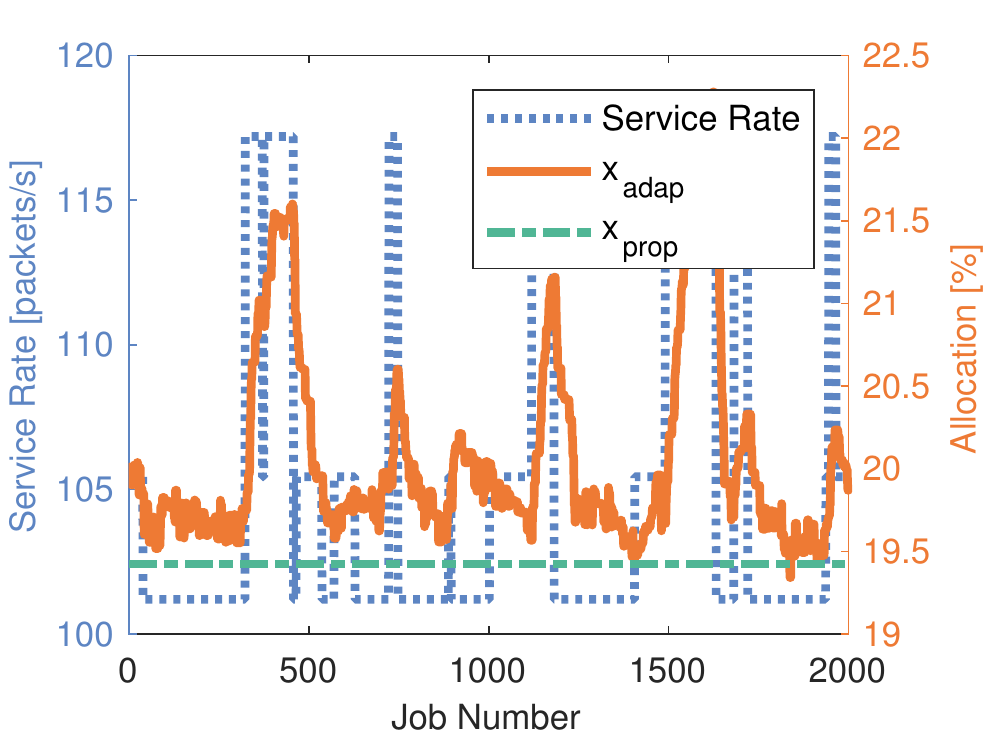}
		\end{subfigure}
		\caption{\label{fig:experiment_3_4}%
		\textbf{Waiting times for \emph{Experiment 1}:}
	We compare a low stress regime \textbf{(Left)} and a high stress regime \textbf{(Middle)}. In both regimes, our adaptive allocation outperforms proportional allocation and batch JSQ.
	Notably the rigidity  of the proportional allocation  causes  large waiting times in the high stress regime.
		\textbf{(Right)} Allocation adaptation 
		on path 1.
	The dotted line shows the  service rate, while the dashed line shows the proportional allocation and the solid line shows the adaptation.
		}
	\end{figure*}

	We consider the setup in Fig.~\ref{fig:collaborative_uploading} with arrivals at the primary device being modeled as a Markov modulated Poisson process (MMPP) to
allow for
burstiness in batch arrivals. This model was shown to be a good candidate for some network traffic \cite{Feldmann1998fitting}. We assume inter-arrival times  are modulated by a three-state Markov chain.
	Further, the batch sizes 
	are sampled from a Poisson distribution with mean $10^2$ to account for varying data sizes to be uploaded. 
%
We assume five heterogeneous paths are available. 
For the service process, we observe just one sample of the service times for each path and each data batch. 

	We evaluate the performance of our adaptive allocation 
	using
	  two experiments. In the first experiment, we   vary the service time distributions to reflect different regimes of stress on the paths, assuming full knowledge of the model parameters. In the second experiment, we do not assume knowledge of the model parameters, but instead infer them. The MC estimate of the subgradient is always based on $10^2$ samples, except for the one sample estimate (OSE) method where only the observed sample is used.  
		We use $10^4$ and $5\cdot 10^3$ simulations for the first and the second experiment, respectively.

 We compare our adaptive allocation method with the proportional allocation because of its observed near-optimality in the intermittent case and in the bound approach  for continuous stream uploading. Note that finding the optimal allocation is computationally expensive, while the proportional allocation is readily found. We also consider a  variant of a queue-aware \quotes{join the shortest queue} (JSQ)
schedule where the  batch is assigned to the path with the shortest queue. Such schedules are known to have good performance. However, for many applications, 
obtaining the queue length information is hard.

\begin{figure}[!b]
\vspace{-.6cm}
\centering
\begin{subfigure}[t]{.24\textwidth}
\includegraphics[width=\textwidth]{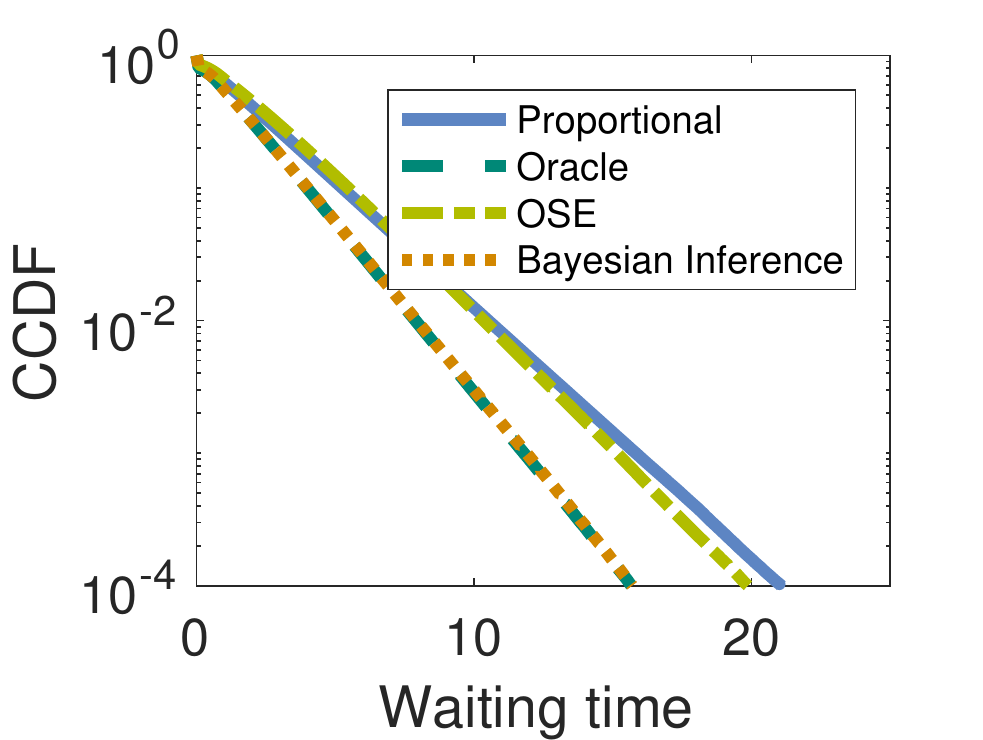}

\end{subfigure}
\hfill
\begin{subfigure}[t]{.24\textwidth}
\includegraphics[width=\textwidth]{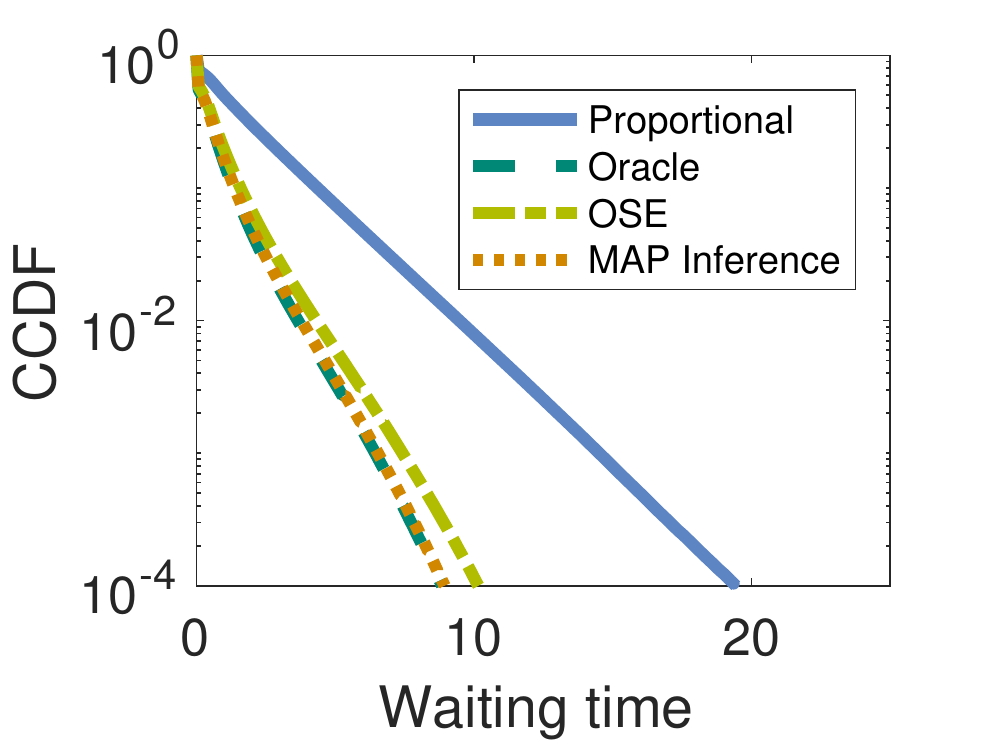}
\end{subfigure}
\caption{\label{fig:exp1_2}%
Waiting time distributions 
for \emph{Experiment 2}. We consider i.i.d  \textbf{(Left)} and Markov modulated \textbf{(Right)} exponentially distributed service times. Evidently the lack of knowledge of the model parameters can be overcome by means of inference, which performs almost as good as the oracle method.
}
\vspace{-3pt}
\end{figure}

		\emph{Experiment 1:}
		Here, we assume Markov modulated  exponentially distributed service times. 
		This captures the service time correlations, \eg,
in time varying wireless channels \cite{MahmoodRJ11}. We use five independent three-state Markov chains to modulate the means of the service times for each chunk on each path.
		For the proportional allocation, we calculate the mean service rate weighted by the stationary probabilities.

We consider two complimentary situations: one, called the \emph{low stress} regime, where the service rates are high such that one path is sufficient to serve incoming batches (ensuring stability), and the other, called the \emph{high stress} regime, where the service rates are low that utilizing all the five paths is \emph{necessary} to ensure stability. In Fig.~\ref{fig:experiment_3_4}, we show  the complementary cumulative distribution function  (CCDF) of the waiting times comparing different allocation strategies. 
This figure shows the benefit of adaptive allocation. Note that the proportional allocation is not adaptive. On the other hand, the batch JSQ under-utilizes parallelization. Our allocation scheme described in Sec.~\ref{subsec:stoch_gradient} ensures adaptiveness while being allocative.
%
%
This benefit of adaptation is prominent, especially in the high stress regime as seen in Fig.~\ref{fig:experiment_3_4} (Middle).
%
Fig.~\ref{fig:experiment_3_4} (Right) also shows how our allocation adapts to the service rate changes. 
Here, the latent state of the Markov chain and the  parameters for the service times are assumed to be known.

\emph{Experiment 2:} In this experiment,  we 
infer the model parameters. In keeping with the setup in Fig.~\ref{fig:adaptive_allocation}, we first assume i.i.d. exponentially distributed service times. Fig.~\ref{fig:exp1_2}~(Left) shows the CCDFs of the waiting times  for different allocation strategies. 
To estimate the subgradient, the oracle  uses the true parameters 
to draw  
samples, 
while the Bayesian inference  draws 
samples from the predictive distribution. 

We further consider the setup in Fig.~\ref{fig:adaptive_allocation} with Markov modulated  exponentially distributed service times  as described in \emph{Experiment 1}. 
The inference method draws samples from the emission distribution using a  MAP 
estimate of the current latent state of the Markov chain for each allocation. The parameters of the Markov modulation of the service times are learned offline using a training sequence. 
In Fig.~\ref{fig:exp1_2}~(Right), we compare our inference-based method with the OSE, 
and an oracle that draws samples from the emission distribution with the true parameters and the true latent states.

From Fig.~\ref{fig:exp1_2}, we see that increasing the number of  samples  for the subgradient estimation leads to smaller tail probabilities, since the subgradient noise decreases. Interestingly, the lack of knowledge of the model parameters does not affect the performance of our adaptive allocation, as the inference method achieves  results comparable to that of the oracle.

%% file: sections/Related_Work.tex
\section{Related Work}
\label{sec:related_work}
The  work in \cite{howe2006rise} anticipated the emergence of crowdsourcing systems.
A recent discussion introduced crowdsourced live event coverage in \cite{Richerzhagen2016Crowd}, where the authors  propose adaptive strategies for collaboratively uploading the most relevant streams. Our analytical treatment is complimentary to  \cite{Richerzhagen2016Crowd}.

The intermittent  uploading scenario is analyzed in \cite{Zhang2011Delay}, where the authors provide an upper bound on the mean delay for the canonical two-path scenario with exponential path delays.
In contrast, we obtain closed-form expressions for the mean delay for $N\ge2$ paths. We also provide tools and examples of the optimization thereof. 

A segment of related work is concerned with the analysis of controlled and uncontrolled Fork-Join
 systems. For uncontrolled, \ie, non-adaptive FJ systems, it is known that exact results are hard to obtain~\cite{Baccelli89}. Exact results are known for the joint workload distribution for only two parallel queues with Poisson arrivals and i.i.d exponential service times~\cite{flatto1984two}.
For  more general scenarios
we resort, \eg, to  bounds on the tail probabilities of the steady-state waiting times for single-stage systems \cite{KhudaBukhsh2017infocom,Rizk2015Sigmetrics,Baccelli89} or multistage systems~\cite{Fidler2016Delay}. They also highlight the benefits of parallelization under high utilization regimes, in agreement with  what we observe in \emph{Experiment 1} in Sec.~\ref{sec:numerical_results}.
The work in \cite{Pedarsani2014} considers controlling FJ systems 
using a gradient descent approach to minimize  queue lengths. They use changes in queue lengths to find an unbiased estimate of the gradient.  
Note that obtaining the 
queue lengths is not straightforward in many applications.
Therefore we  infer the service time distributions 
to adapt our allocation using gradient descent to minimize the expected waiting time.
In~\cite{Sun2017near}, the authors investigate scheduling of batch jobs in  systems with multiple servers. As opposed to our setup, they assume a queuing model without synchronization constraint and  only consider a special class of i.i.d. distributed service times. For our adaptive allocation method, we do not make any independence assumption.                          
%
%
%


Redundancy techniques have grown in popularity over the years as a means to decrease latency. 
In \cite{Vulimiri:2013:LLV}, the authors study the trade-off between the latency reduction attained by redundancy and the corresponding overhead. Based on empirical results, they argue that redundancy can be effective 
in a large class of applications. 
 The work in \cite{Joshi:2017:ERT} is close to ours. The authors model a cloud computing scenario as a Fork-Join  system with identical servers and analyze different redundancy techniques,  which are akin to our  $(N,r)$-allocations in the intermittent uploading case,  with a view to reducing latency in a cost-efficient manner. The authors find that the log-concavity of the task service times decides the success of redundancy techniques. 
Their approach is complimentary to our adaptive allocation  with heterogeneous Markov modulated servers. 

The allocation problem  in the continuous stream uploading case can be seen as a type of load-balancing problem, 
however, with the mean waiting time as the objective function. 
A programming model for the allocation of continous streams over multiple paths was introduced in \cite{Frommgen2017programming}. In a recent work \cite{shah2017delay}, the authors consider storage and delivery of large files in data-centers, where 
files are first erasure-coded  and then stored in a subset of the available servers. They  compare the performance of water-filling and  batch sampling  as dynamic load-balancing policies
and provide 
computable performance bounds. In contrast, we do not restrict ourselves to a rigid allocation strategy and adapt our allocation dynamically. 

%% file: sections/Discussion.tex
\section{Conclusion}
\label{sec:discussion}
In this work, we optimize allocation and replication strategies in collaborative uploading scenarios.  
We differentiate between
%
intermittent  and continuous stream uploading, based on the system's queuing behavior. In the first case (no queuing),
we unify the notions of allocation and replication, and
provide closed-form expressions for the mean upload latency. We use our exact formulation for the intermittent uploading case to derive optimal allocation and replication strategies.  

We pose the continuous  stream uploading case as a Fork-Join queuing model with varying burstiness of the data traffic to be uploaded, and of the paths' service.
Thereby we propose an adaptive allocation scheme, based on statistical inference of the properties of the paths' latencies. 
We sequentially  minimize a notion of the expected waiting time,  ensuring a bounded regret.
We show the effectiveness of our adaptive approach compared to proportional allocation and batch JSQ allocation.
The lack of knowledge of the model parameters does not affect the performance of our adaptive allocation, as the inference methods are able to achieve results comparable to those of an oracle with full system knowledge.

%% file: sections/Appendix.tex

\section{}
\label{sec:AppendixA}

\subsection{Moments of order statistics}
\label{subsec:order_stat}
Let $X_1,X_2,\ldots,X_N$ be independent positive-valued random variables with \emph{absolutely continuous} CDFs $F_1,F_2,\ldots, F_N$. 
Let the corresponding order statistics be $Y_1\leq Y_2\leq \ldots \leq Y_N$.
Write $\vec{F} \defeq (F_1,F_2,\ldots, F_N)\trans$ and $\vec{1}-\vec{F} \defeq (1-F_1,1-F_2,\ldots,1- F_N)\trans$.  
The distribution of the $r$-th order statistic can be elegantly
written  in terms of certain permanents  as
   \cite[Theorem 4.1]{bapat1989perm}, 
\begin{align}
\probOf{Y_r \le y} =&{} \sum_{i=r}^{N} \frac{1}{i! (N-i)!} \Permanent{ \Big[ \frac{\vec{F}(y)}{i} \frac{\vec{1}-\vec{F} (y)}{N-i} \Big]  } \eqcomma
\label{eq:distoforderstat}
\end{align}
where $\Big[ \frac{\vec{F}(y)}{i} \frac{\vec{1}-\vec{F} (y)}{N-i} \Big] $ denotes the matrix whose first~$i$ columns are $\vec{F}(y)$ and the last~$N-i$ columns are~$\vec{1}-\vec{F}(y)$,  $\Permanent{A} \defeq \sum_{\sigma \in \allpermutations{N}}  \prod_{i=1}^{N} a_{i, \sigma(i)} $ denotes the permanent of an $N \times N$ real matrix $A = ((a_{i,j}))_{i,j \in \setN{N}}$,  and $\allpermutations{N}$ denote the class of all permutations of $\setN{N}$. Using \eqref{eq:distoforderstat}, we  derive the expected values 
of the order statistics \cite{bapat1989perm,Barakat2004Moments}\ifthenelse{\boolean{longVersion}}{}{ (proof given in \cite{KhudaBukhsh2017TechReport})}. 
\begin{myRemark}
For $ r \in \setN{N}$, the mean  of $Y_r$ can be conveniently written in terms of $\myMuoperator{}{}$-operators given by
\begin{align*}
 \Eof{ Y_r  } = \myMuoperator{r}{\vec{F}} \defeq \sum_{j = N-r+1}^{N} (-1)^{j- (N-r-1)} \binom{j-1}{N-r} \myDoperator{j}{\vec{F}}  \, \eqcomma
\end{align*}
 where the $ \myDoperator{j}$-operators,  
  for $j \in \setN{N}$, are defined as
  \begin{align}
  \myDoperator{j}{\vec{F}}  \defeq \sum_{ \substack{ S  \in \{ A   \subseteq \setN{N} : \cardinality{A}=j  \} }  }  \int_{0}^{\infty}  \big( \prod_{i \in S} (1- F_i(x))  \big) \, \mathrm{d}x \, \eqstop
  \label{eq:myDoperator}
  \end{align}
\label{remark:moment_order_stat}
\end{myRemark}

\ifthenelse{\boolean{longVersion}}{
\begin{proof}[Proof of Remark~\ref{remark:moment_order_stat}]
  The proof follows  from \cite{bapat1989perm,Barakat2004Moments}. However, for the sake of completeness, we furnish a brief sketch here. Define, for $r \in \setN{N}$,
  \begin{align}
    H_r(y) \defeq \probOf{Y_r \le y} \eqcomma
  \end{align}
  where $\probOf{Y_r \le y}$ is given in \eqref{eq:distoforderstat}. Then, the mean can be obtained by performing the following integral
  \begin{align*}
     \Eof{ Y_r  } = \int_0^\infty \left(1- H_r(y) \right)\, \mathrm{d}y \eqstop
  \end{align*}

Observe that, we can derive the following recursion relation from \eqref{eq:distoforderstat}, for $r\ge 2$,
\begin{align}
H_{r-1}(y) &= H_r(y) +    \frac{1}{ (r-1)! (N-r+1)!   } \Permanent{  \Big[ \frac{\vec{F}(y)}{r-1} \frac{\vec{1}-\vec{F} (y)}{N-r+1} \Big]   }  \eqcomma
\label{eq:h_recurrence}
\end{align}
where the permanent of a real $N\times N$ matrix $A \defeq ((a_{i,j}))_{i,j \in \setN{N}}$ is given by
\begin{align*}
  \Permanent{A} \defeq \sum_{\sigma \in \allpermutations{N}}  \prod_{i=1}^{N} a_{i, \sigma(i)} \eqcomma
\end{align*}
and $\allpermutations{N}$ denote the class of all permutations of $\setN{N}$. Plugging in the definition of the permanent, we rewrite \eqref{eq:h_recurrence} as
\begin{align*}
H_{r-1}(y) &= H_r(y) +  \frac{1}{ (r-1)! (N-r+1)!   } \sum_{\sigma \in \allpermutations{N}}  \prod_{i=1}^{N} a_{i, \sigma(i)} \left(y \right)  \eqcomma
\end{align*}
where
\begin{align}
  a_{i, \sigma(i)} \left(y \right) =&  \left\{ \begin{array}{ll}
    F_i(y) & \mbox{ if }  1 \le \sigma(i) \le r-1  \eqcomma  \\
    1-  F_i(y) & \mbox{ if }  r \le \sigma(i) \le N \eqstop
\end{array}
\right.
\end{align}
Rearranging the terms in the recurrence relation, we get
\begin{align*}
  1-  H_r(y) =&  1- H_{r-1}(y) \\
  &{} +  \frac{1}{ (r-1)! (N-r+1)!   } \sum_{\sigma \in \allpermutations{N}}  \prod_{i=1}^{N} a_{i, \sigma(i)} \left(y \right)  \eqstop
\end{align*}
Integrating both sides and using the $ \myMuoperator{}{}$-operators, we get
\begin{align*}
\myMuoperator{r}{  \vec{F}  }   = &\myMuoperator{r-1}{ \vec{F}   } \\
 & {} + \frac{1}{ (r-1)! (N-r+1)!   } \sum_{\sigma \in \allpermutations{N}}    \int_0^\infty   \prod_{i=1}^{N}   a_{i, \sigma(i)} \left(y \right)   \, \mathrm{d}y  \\
 = & \myMuoperator{r-1}{ \vec{F}   }  + K_r \vec{F} \eqcomma
\end{align*}
where the operator $K_r$ is given by
\begin{align*}
  K_r \vec{F}  \defeq & \frac{1}{ (r-1)! (N-r+1)!   } \sum_{\sigma \in \allpermutations{N}}    \int_0^\infty   \prod_{i=1}^{N}   a_{i, \sigma(i)} \left(y \right)   \, \mathrm{d}y  \eqstop
\end{align*}
Note that there are $r-1$ terms involving $F_i(y)$ and $N-r+1$ terms involving $1- F_i(y)$ in the product, for each permutation $\sigma \in \allpermutations{N}$. Therefore, we have
\begin{align*}
  & K_r \vec{F}
  = 
  & {}  \sum_{ S  \in \{A   \subseteq \setN{N}: \cardinality{A}= r-1 \}   }  \int_0^\infty \left(  \prod_{j \in S}  F_j(y)  \right) \left( \prod_{j \in S^c}  (1- F_j(y))   \right) \, \mathrm{d}y \eqstop
\end{align*}
Let us rewrite $K_r$-operators in the following way to get an identity
\begin{align}
  K_r \vec{F}   \equiv & {}  \sum_{j =1 }^{r}  (-1)^{j-1}  c(j, r, N)  \,  \myDoperator{N-r+j}{ \vec{F}  }
   \eqcomma
\label{eq:K_operator_identity}
\end{align}
where $c(j,r,N)$'s are  suitable counting coefficients so that the above identity holds true with $\myDoperator{}{}$-operators defined by
\begin{align*}
    \myDoperator{j}{\vec{F}}  \defeq \sum_{ \substack{ S  \in \{ A   \subseteq \setN{N} : \cardinality{A}=j  \} }  }  \int_{0}^{\infty}  \big( \prod_{i \in S} (1- F_i(x))  \big) \, \mathrm{d}x \, \eqstop
\end{align*}
Notice that the number of terms under the summation over $ S \subseteq \setN{N}$ with  $ \cardinality{S}= r-1    $ is $\binom{N}{r-1}$, while that under the summation over $ S \subseteq \setN{N}$ with $ \cardinality{S}=N-r+j  $ appearing in the computation of $ \myDoperator{N-r+j}{ \vec{F}  }   $ is $\binom{N  }{  N-r+j }$. Therefore, by applying multiplication principle of combinatorial analysis, the counting coefficients $c(j,r,N)$ must satisfy
\begin{align*}
  \binom{N}{r-1} \binom{r-1}{j-1} = c(j,r,N) \binom{N}{N-r+j} \eqcomma
\end{align*}
in order for the above identity  in \eqref{eq:K_operator_identity} to hold true (see \cite{Barakat2004Moments}). Therefore, we get
\begin{align}
  c(j,r,N)  = & \binom{N-r+j}{ j-1 } \eqcomma
\end{align}
and we get the following recursion relation, for $2 \le r \le N$,
\begin{align}
  \myMuoperator{r}{  \vec{F}  }   = &\myMuoperator{r-1}{ \vec{F}   }  +  \sum_{j =1 }^{r}  (-1)^{j-1}  \binom{N-r+j}{ j-1 }  \,  \myDoperator{N-r+j}{ \vec{F}  }  \eqstop
  \label{eq:mu_recursion}
\end{align}

Observe that $\myMuoperator{1}{  \vec{F} }  = \myDoperator{N}{    \vec{F}}$ and $\myMuoperator{2}{ \vec{F} } = \myDoperator{N-1}{ \vec{F}} - (N-1) \myDoperator{N}{ \vec{F}}$. Thereby from \eqref{eq:mu_recursion}, the  claim
\begin{align}
  \myMuoperator{r}{\vec{F}} = \sum_{j = N-r+1}^{N} (-1)^{j- (N-r-1)} \binom{j-1}{N-r} \myDoperator{j}{\vec{F}}
\end{align}
follows by induction on $r$. The induction is proved in \cite{Barakat2004Moments} and we do not repeat it here. This completes the proof.

\end{proof}
}{}

\subsection{General $(N,r)$-strategy}

\begin{myExample}[Example of $(N,r)$-strategy]
  \label{example:gen_n_r_strategy}
	Suppose we have three  paths with exponential delays with parameters~$\lambda_1, \lambda_2$ and $\lambda_3$. Define, for $i=1,2,3$,  $p_{ij}=\frac{\lambda_i}{\lambda_i+\lambda_j}, \, q_{ij}=1-p_{i,j}, \, r_{ij}= \frac{1}{\lambda_i+\lambda_j}$ and $p_{123}^{(i)}= \frac{\lambda_i}{\lambda_1+\lambda_2+\lambda_3}, \, r_{123}= \frac{1}{\lambda_1+\lambda_2+\lambda_3}$.   
  The mean upload latency corresponding to a  $(3,1)$-allocation (replication) $\vec{k} =(k_1,k_2,k_3) \in \genAllocation{3}{1}{K}$  is $\myMuoperator{1}{ \vec{F}^{(\vec{k})} } =\eta_1(\vec{k})$,  and is given by

  \begin{align*}
	& r_{123} \sum_{n_1=0}^{k_1-1}\sum_{n_2=0}^{k_2-1} \sum_{n_3=0}^{k_3-1} \frac{  (n_1+n_2+n_3)!  }{n_1! n_2! n_3!} \left( p_{123}^{(1)} \right)^{n_1}
  \left( p_{123}^{(2)} \right) ^{n_2} \left( p_{123}^{(3)} \right)^{n_3} \eqstop 
\end{align*}
  %

For a $(3,2)$-allocation $\vec{k}  \in \genAllocation{3}{2}{K}$, the mean upload latency,  $ \myMuoperator{2}{ \vec{F}^{(\vec{k})} } = \eta_2 ( \vec{k} ) =\myDoperator{2}{\vec{F}^{(\vec{k})} } - 2 \myDoperator{3}{\vec{F}^{(\vec{k})}}  $, is given by

\begin{align*}
& r_{12} \sum_{n_1=0}^{k_1-1}\sum_{n_2=0}^{k_2-1}   \frac{  (n_1+n_2)!  }{n_1! n_2!} p_{12}^{n_1}q_{12}^{n_2}
 + r_{23} \sum_{n_2=0}^{k_2-1}\sum_{n_3=0}^{k_3-1} \frac{  (n_2+n_3)!  }{n_2! n_3!} p_{23}^{n_2}q_{23}^{n_3} \nonumber \\
& + r_{31} \sum_{n_3=0}^{k_3-1}\sum_{n_1=0}^{k_1-1} \frac{  (n_3+n_1)!  }{n_3! n_1!} p_{31}^{n_3} q_{31}^{n_1}
 -2  \eta_1(\vec{k})  \eqstop 
\end{align*}
%
%
Note that a $(3,3)$-allocation corresponds to simple allocation (see Sec.~\ref{subsec:multipath_exp}). The derivations 
are provided in \ifthenelse{\boolean{longVersion}}{Appendix~\ref{sec:AppendixB}}{\cite{KhudaBukhsh2017TechReport}}.
\end{myExample}

\ifthenelse{\boolean{longVersion}}{
\subsection{The bound approach}

\begin{myDefinition}[Steady-state waiting times in a Fork-Join queuing system]
	For simplicity, we assume we have a fixed allocation~$\vec{k}=(k_1,k_2,\ldots,k_N) \in \diophantine{N}{K}$ and the service times are independent. Let $t_i$ denote the inter-arrival time between the $i$-th and the $i+1$-th jobs.  Let $S_{i,j}^{(k_i)}$ denote the amount of time taken (service time) by server (path)~$i$ to transport a chunk of size~$k_i$ of the $j$-th job (file), for $i \in \setN{N}, j \in \setOfNaturals$. To draw a parallel, $S_{i,j} ^{(k_i)}$'s are independent copies of $S_{i}^{(k_i)}$ from Sec.~\ref{subsec:single-file}. Then, formally the waiting time of the $j$-th job is defined as $	\max\{ 0, \sup_{k \in \setN{j-1}} \{  \sup_{l \in \setN{N}} \{ \sum_{i=1}^{k} S_{l,j-i} ^{(k_l)}  -\sum_{i=1}^{k} t_{j-i}  \} \}   \} \eqcomma $ for $j>1$ and $0$ for $j=1$ (see \cite{Rizk2015Sigmetrics,KhudaBukhsh2017infocom,KhudaBukhsh2016GenPE}), from which we have the following steady state representation of the waiting time $W$:
	\begin{align}
	W  \disteq  \sup_{k \in \setOfNonnegativeIntegers}  \sup_{l \in \setN{N}} \{ \sum_{j=1}^{k} S_{l,j} ^{(k_l)}  -\sum_{j=1}^{k} t_{j}  \}   \eqcomma
	\label{eq:steady_state}
	\end{align}
	where $\disteq$ denotes equality in distribution. Our goal is to find an allocation that ensures as little waiting times as possible.

	It is infeasible to find closed form expression for the distribution of the steady state waiting times under general settings. However, one can find reasonably good bounds. Following \cite{KhudaBukhsh2017infocom}, we get
	\begin{align}
	 \probOf{ W \geq \sigma }  \leq {}& \myExp{ - \tilde{\theta}  \sigma  }  \sum_{i \in \setN{N} } \myExp{ -  ( \theta_i -  \tilde{\theta}   )\sigma }  \eqcomma
	\end{align}
	where $\theta_i$ is the positive solution of $ \Eof{ \myExp{  S_{i,1} ^{(k_i)} -  t_{1}    }     }  =1$
	for $i \in \setN{N}$ and $\tilde{\theta} \defeq \min_{i \in \setN{N} } \theta_i$.
\end{myDefinition}

\begin{myExample}[The canonical two-path scenario]
  \label{example:decay_rate_exp}
  Suppose we have two heterogeneous paths with exponential delays. Let the rates of the exponential distributions be $\lambda_1$ and $\lambda_2$. Also, suppose the arrival process is renewal with exponentially distributed inter-arrival times. Let the rate of the inter-arrival distribution be $\lambda_{\text{a}}$.

  Consider an allocation~$\vec{k} \defeq (k_1,k_2) \in \diophantine{2}{K}$. Then,  $ S_{1,1} ^{(k_1)}$ and $ S_{2,1} ^{(k_2)}$ are gamma distributed with shape and scale  parameters $(k_1, \lambda_1)$ and $(k_2, \lambda_2)$ respectively. Then $\theta_1$ and $\theta_2$ are the solutions of the following equations
  \begin{align*}
    \left( 1- \frac{\theta_1}{\lambda_1}  \right)^{-k_1} \left(1+  \frac{\theta_1}{\lambda_{\text{a}}}  \right) &=1 \eqcomma    \\
        \left( 1- \frac{\theta_2}{\lambda_2}  \right)^{-k_2} \left(1+  \frac{\theta_2}{\lambda_{\text{a}}}  \right) &=1 \eqstop
  \end{align*}
  Solving the above two equations, we get the effective decay rate as
  \begin{align}
    \tilde{\theta} &= \min  \left(  \theta_1, \theta_2  \right) \eqstop
  \end{align}

In Fig.~\ref{fig:theta_opt}, we show how the effective decay rate depends on the data size~$K$.

\end{myExample}

}{}

%

%% file: sections/Appendix_B.tex
\section{}
\label{sec:AppendixB}

\begin{proof}[Proof of    $\psi(k_1,k_2)  \gtreqless \psi(k_1+1,k_2-1)  \iff \frac{ I_{ p } (k_1, k_2  )     }{I_{ 1-p } (k_2-1, k_1+1  ) }  \gtreqless \frac{\lambda_2}{\lambda_1} $ 
	]
	Write $p\defeq \frac{\lambda_1}{\lambda_1+\lambda_2}$ and $q\defeq \frac{\lambda_2}{\lambda_1+\lambda_2}$.
\begin{align*}
\psi(k_1,k_2) =& \frac{k_1}{\lambda_1}+ \frac{k_2}{\lambda_2}
 - \frac{1}{\lambda_1+\lambda_2} \sum_{n_1=0}^{k_1-1}\sum_{n_2=0}^{k_2-1} \binom{n_1+n_2}{n_1} p^{n_1} q^{n_2} \\
=&   \frac{k_1}{\lambda_1}+ \frac{K-k_1}{\lambda_2}  - \frac{1}{\lambda_1+\lambda_2} \sum_{n_1=0}^{k_1-1}\sum_{n_2=0}^{K-k_1-1} \binom{n_1+n_2}{n_1} p^{n_1} q^{n_2} \eqstop
\end{align*}
Now,

\begin{align*}
&\psi(k_1,k_2)  - \psi(k_1+1,k_2-1) \\
= & \frac{k_1}{\lambda_1}+ \frac{K-k_1}{\lambda_2}
 - \frac{1}{\lambda_1+\lambda_2} \sum_{n_1=0}^{k_1-1}\sum_{n_2=0}^{K-k_1-1} \binom{n_1+n_2}{n_1} p^{n_1} q^{n_2} \\
& {} - \big[ \frac{k_1+1}{\lambda_1}+ \frac{K-k_1-1}{\lambda_2} \\
&{} - \frac{1}{\lambda_1+\lambda_2} \sum_{n_1=0}^{k_1}\sum_{n_2=0}^{K-k_1-2} \binom{n_1+n_2}{n_1} p^{n_1} q^{n_2} \big] \\
= & \big( \frac{1}{\lambda_2}- \frac{1}{\lambda_1} \big)
  - \frac{1}{\lambda_1 +\lambda_2} \big[  \sum_{n_1=0}^{k_1-1}\sum_{n_2=0}^{K-k_1-1} \binom{n_1+n_2}{n_1} p^{n_1} q^{n_2} \\
  & {}  -  \sum_{n_1=0}^{k_1}\sum_{n_2=0}^{K-k_1-2} \binom{n_1+n_2}{n_1} p^{n_1} q^{n_2}   \big] \eqstop
\end{align*}
Simplifying further,
\begin{align*}
& \sum_{n_1=0}^{k_1-1}\sum_{n_2=0}^{K-k_1-1} \binom{n_1+n_2}{n_1} p^{n_1} q^{n_2}
 -  \sum_{n_1=0}^{k_1}\sum_{n_2=0}^{K-k_1-2} \binom{n_1+n_2}{n_1} p^{n_1} q^{n_2}   \\
 =&  \sum_{n_1=0}^{k_1-1} \big[ \sum_{n_2=0}^{K-k_1-2} \binom{n_1+n_2}{n_1} p^{n_1} q^{n_2}  + \binom{n_1+K- k_1-1}{n_1} p^{n_1} q^{K- k_1-1} \big] \\
 & {} - \big[   \sum_{n_1=0}^{k_1-1}\sum_{n_2=0}^{K-k_1-2} \binom{n_1+n_2}{n_1} p^{n_1} q^{n_2} +  \sum_{n_2=0}^{K-k_1-2} \binom{k_1+n_2}{k_1} p^{k_1} q^{n_2}   \big] \\
 =&  \sum_{n_1=0}^{k_1-1} \binom{n_1+K- k_1-1}{n_1} p^{n_1} q^{K- k_1-1}  - \sum_{n_2=0}^{K-k_1-2} \binom{k_1+n_2}{k_1} p^{k_1} q^{n_2} \\
 =& \frac{1}{q} \sum_{n_1=0}^{k_1-1} \binom{n_1+K- k_1-1}{n_1} p^{n_1} q^{K- k_1}  - \frac{1}{p}  \sum_{n_2=0}^{K-k_1-2} \binom{k_1+n_2}{k_1} p^{k_1+1} q^{n_2} \\
 =&  \frac{1}{q}  F(k_1-1; K-k_1,p) -  \frac{1}{p}  F(K-k_1-2; k_1+1,p) \\
 =& \frac{1}{q} I_q(K-k_1,k_1 ) -  \frac{1}{p} I_p( k_1+1, K- k_1-1) \eqcomma
\end{align*}
where $F(.; n, s)$ is the cumulative distribution function of a negative binomial distribution with parameters $n$ and $s$ (denoted as $\NegBin{n}{s}  $) and $I_{x}(a,b)$ is the regularized $\beta$-function given by
\begin{align*}
I_{x}(a,b) \defeq \frac{   \int_{0}^{x} t^{a-1}   (1-t)^{b-1}  \, dt  }{   \int_{0}^{  1 } t^{a-1}   (1-t)^{b-1}  \, \mathrm{d}t   } \eqstop
\end{align*}
Therefore we have
\begin{align*}
& \psi(k_1,k_2)   -  \psi(k_1+1,k_2-1) \\
= &  \big( \frac{1}{\lambda_2}- \frac{1}{\lambda_1} \big)  - \frac{1}{\lambda_1 +\lambda_2} \big[ \frac{1}{q} I_q(K-k_1,k_1 ) \\
&{}
 -  \frac{1}{p} I_p( k_1+1, K- k_1-1)  \big] \\
=&  \frac{1}{\lambda_2} \big(1-I_q(K-k_1,k_1 ) \big)
  - \frac{1}{\lambda_1} \big( 1-  I_p( k_1+1, K- k_1-1)  \big) \\
  =& \frac{1}{\lambda_2} I_p(k_1,K-k_1 ) - \frac{1}{\lambda_1} I_q( K- k_1-1, k_1+1 )  \eqcomma
\end{align*}
whence we get
\begin{align*}
\psi(k_1,k_2)   \gtreqqless & {}  \psi(k_1+1,k_2-1) \\
\iff \frac{1}{\lambda_2} I_p(k_1,K-k_1 )  \gtreqqless & {}   \frac{1}{\lambda_1} I_q( K- k_1-1, k_1+1 ) \\
\iff \frac{    I_p(k_1,K-k_1 )  }{I_q( K- k_1-1, k_1+1 )   } \gtreqqless \frac{\lambda_2}{\lambda_1} \eqstop
\end{align*}
This completes the proof.

This allows finding the optimal allocation $\vec{k}_{\text{opt}}$
in an iterative fashion.  
Given $\lambda_1 < \lambda_2$, and we sequentially check $(0,K), (1,K-1), (2, K-2), \ldots$  and so on 
as long as the ratio of the two regularized $\beta$-functions is greater than $\lambda_2/\lambda_1$.
The objective function $\psi$ is monotonically decreasing in its first argument $k_1$ in this range. The optimal choice is  the last allocation in this sequence when the ratio of the regularized $\beta$-functions is greater than or equal to  $\lambda_2/\lambda_1$, beyond this point 
$\psi$ is again monotonically increasing in its first argument $k_1$. See Fig.~\ref{fig:two-path}.
If $\lambda_1 > \lambda_2$, we interchange (relabel) the paths and proceed as before. 
Since the optimal allocation is $(K/2,K/2)$ (or the nearest integers depending on whether $K$ is even or odd) when 
$\lambda_1=\lambda_2$.

\end{proof}

\begin{proof}[Proof of optimality when $\lambda_1=\lambda_2$]

  Suppose $\lambda_1 = \lambda_2=\lambda$,
    \begin{align*}
    \psi(k_1,k_2) =& \frac{K}{\lambda}
    - \frac{1}{2\lambda} \sum_{n_1=0}^{k_1-1}\sum_{n_2=0}^{k_2-1} \binom{n_1+n_2}{n_1} \big(\frac{1}{2}\big)^{n_1+n_2} \eqstop
    \end{align*}

For natural numbers $c, m$ with $c>m$, define
\begin{align*}
G(m) =&  \sum_{n_1=0}^{m}\sum_{n_2=0}^{c-m} \binom{n_1+n_2}{n_1} \big(\frac{1}{2}\big)^{n_1+n_2} \eqstop
\end{align*}
Then
\begin{align*}
G(m+1)-G(m) =& \sum_{n_2=0}^{c-m-1} \binom{m+1+n_2}{m+1} \big(\frac{1}{2}\big)^{m+1+n_2}\\
&{} - \sum_{n_2=0}^{m} \binom{c-m+n_2}{c-m} \big(\frac{1}{2}\big)^{c-m+n_2} \eqstop
\end{align*}
Comparing the last summands under two summations,
\begin{align*}
\binom{c}{m+1} \big(\frac{1}{2}\big)^{c} \gtreqless {}& \binom{c}{c-m} \big(\frac{1}{2}\big)^{c} \\
\iff& \frac{c-m}{m+1} \gtreqless 1\\
\iff& 2m \lesseqgtr c-1
\end{align*}
Further,
\begin{align*}
\frac{c-m}{m+1} \ge 1
\implies \binom{c-i}{m+1} \big(\frac{1}{2}\big)^{c-i} \gtreqless \binom{c-i}{c-m} \big(\frac{1}{2}\big)^{c-i}
\end{align*}
Summing over $i=0,1,...,\min(c-m-1,m)$
\begin{align*}
2m \le c-1
\implies G(m+1) \ge G(m)
\end{align*}
The converse can be proved using similar arguments. Therefore, $G(m)$ attains maxima at $\floor{\frac{c+1}{2}}$. Consequently, $\psi$ is minimum when $n_1-1=\floor{\frac{k_1-1+k_2-1+1}{2}}$ i.e., $n_1 = \floor{\frac{K+1}{2}}$. This completes the proof.

\end{proof}

\begin{proof}[Derivation of $N$-path case with exponential delays] 
Suppose the $i$-th path has an exponential delay with rate $\lambda_i$ (\ie, $D_{i,j}$'s follow an exponential distribution with mean $1/\lambda_i$, for $i \in \setN{N}$). Let $\vec{k} = (k_1,k_2,\ldots,k_N) \in \diophantine{N}{K}  $ be our allocation. Then,  the end-to-end delay can be expressed as~$D \defeq \max ( D_1^{(k_1)}, D_2^{(k_2)}, \ldots, D_N^{(k_N)}  )$ where $D_i^{(k_i)} \defeq \sum_{j=1}^{k_i} D_{i,j}$. Note that $D_i^{(k_i)}$ follows a gamma distribution with parameters $k_i$ and $ \lambda_i$ (which is the same as an Erlang distribution in this case). Therefore, the cumulative distribution function $ F_i^{(k_i)} $ of $D_i^{(k_i)}$ is given by
\begin{align}
F_i^{(k_i)}(x) = 1- \sum_{m = 0 }^{k_i -1 } e^{- \lambda_i x} \frac{  (\lambda_i x)^m }{m!} \; \text{ for } i \in \setN{N} \eqstop
\end{align}
Stacking into a column vector $\vec{F}^{(\vec{k})} \defeq (F_1^{(k_1)},F_2^{(k_2)},\ldots, F_N^{(k_N)}  )$, and following Remark~\ref{remark:moment_order_stat}, we find explicitly
\begin{align}
\psi(\vec{k})  = \myMuoperator{n}{  \vec{F}^{(\vec{k})} } =  \sum_{j = 1}^{n} (-1)^{j +1 } \myDoperator{j}{  \vec{F}^{(\vec{k})} } \eqcomma
\end{align}
where $\myDoperator{j}{}$'s are as defined in  \eqref{eq:myDoperator} of  Remark~\ref{remark:moment_order_stat}.  To find it explicitly, please note that
\begin{align*}
\myDoperator{j}{  \vec{F}^{(\vec{k})} } =&{}  \sum_{ \substack{ S \in \{A  \subseteq \setN{N} : \cardinality{A}=j   \} }  }  \int_{0}^{\infty} \big[ \prod_{i \in S} (1- F_i^{(k_i)}(x))  \big] \, \mathrm{d}x \\
= &{}  \sum_{ \substack{ S \in  \{A \subseteq \setN{N} : \cardinality{A}=j  \} }  }  \int_{0}^{\infty} \big[  \prod_{i \in S} \sum_{m_i = 0 }^{k_i -1 } e^{- \lambda_i x} \frac{  (\lambda_i x)^{m_i} }{{m_i}!}   \big] \, \mathrm{d}x \eqstop
\end{align*}
Using
\begin{align}
\int_{0}^{\infty }  e^{ax} x^{b-1}\, \mathrm{d}x = \frac{  \Gamma(b)  }{ a^b  } \eqcomma
\end{align}
and rearranging terms, we get
\begin{align*}
\psi(\vec{k}) =& \sum_{\substack{ S   \in \{ A \subseteq \setN{N} : \\ A \neq \phi  \} } }  (-1)^{\cardinality{S} +1} \big[
\sum_{ \substack{ n_i \in \setN{k_i-1} \cup \{0\} : \\  i \in S } }  \big( \prod_{i \in S}  \frac{ \lambda_i^{n_i} }{n_i !} \big) \nonumber \\
& \times \frac{  \Gamma( \sum_{i \in S} n_i +1  )   }{  (  \sum_{i \in S} \lambda_i  )^{\sum_{i \in S} n_i +1   } } \big] \, \eqstop
\end{align*}
This completes the derivation. Please see \cite{bapat1989perm,Barakat2004Moments} for more on this and other similar examples.
\end{proof}

\begin{proof}[Derivation of   an upper bound on the mean upload latency
	]

We have
\begin{align}
\psi(\vec{k}) \defeq &{} \Eof{ \max ( D_1^{(k_1)}, D_2^{(k_N)}, \ldots, D_N^{(k_N)}  )  }  \eqstop
\end{align}

Following   \cite{Boucheron2013Concentration},
\begin{align*}
 &\myExp{y  \Eof{ \max ( D_1^{(k_1)}, D_2^{(k_N)}, \ldots, D_N^{(k_N)}  )  } } \\
&  \le {} \Eof{ \myExp{ y \max ( D_1^{(k_1)}, D_2^{(k_N)}, \ldots, D_N^{(k_N)}  )   }   }  \\
  &\le {} \sum_{i \in \setN{N}  }  \Eof{ \myExp{  y D_i^{(k_i)} }   } \\
 &=  {}  \sum_{i \in \setN{N}  } \kappa_i (y) \eqcomma
 \end{align*}
where $\kappa_i(y) \defeq  \Eof{ \myExp{  y D_i^{(k_i)} }   } $. Optimizing over $y$, we get
\begin{align}
\psi(k_1,k_2,\ldots,k_N)
\leq & {} \inf_{y \in  \cap_{ i \in \setN{N} } \effectiveDomain{   \kappa_i}  } \frac{1}{y} \log( \sum_{i \in \setN{N}  } \kappa_i (y)  )  \eqstop
\end{align}

\end{proof}


\begin{proof}[Derivations for Example~\ref{example:gen_n_r_strategy}]
	\begin{align}
& \eta_1(\vec{k}) \nonumber  \\
=& \myMuoperator{1}{ \vec{F}^{(\vec{k})}  }  \nonumber  \\
=& \myDoperator{3}{\vec{F}^{(\vec{k})} } \nonumber \\
=& \int_{0}^{\infty} (1- F_1^{(k_1)}(x))  (1- F_2^{(k_2)}(x))  (1- F_3^{(k_3)}(x)) \, \mathrm{d}x \nonumber \\
=& \int_{0}^{\infty} \big(  \sum_{n_1 = 0 }^{k_1-1 } e^{- \lambda_1 x} \frac{  (\lambda_1 x)^n_1 }{n_1!}  \big) \big(  \sum_{n_2 = 0 }^{k_2-1 } e^{- \lambda_2 x} \frac{  (\lambda_2 x)^n_2 }{n_2!}  \big) \nonumber \\
& {} \times  \big(  \sum_{n_3 = 0 }^{k_3-1 } e^{- \lambda_3 x} \frac{  (\lambda_3 x)^n_3 }{n_3!}  \big)  \, \mathrm{d}x \nonumber  \\
=&  \frac{1}{\lambda_1+\lambda_2+\lambda_3} \sum_{n_1=0}^{k_1-1}\sum_{n_2=0}^{k_2-1} \sum_{n_3=0}^{k_3-1} \frac{  (n_1+n_2+n_3)!  }{n_1! n_2! n_3!} \big(\frac{\lambda_1}{\lambda_1+\lambda_2+\lambda_3}\big)^{n_1} \nonumber \\
& \times   \big(\frac{\lambda_2}{\lambda_1+\lambda_2+\lambda_3}\big)^{n_2}  \big(\frac{\lambda_3}{\lambda_1+\lambda_2+\lambda_3}\big)^{n_3} \eqcomma
\end{align}
using
\begin{align*}
\int_{0}^{\infty }  e^{ax} x^{b-1}\, \mathrm{d}x = \frac{  \Gamma(b)  }{ a^b  } \eqstop
\end{align*}

\begin{align}
& \eta_2 ( \vec{k} ) \nonumber \\
=& {}  \myMuoperator{2}{\vec{F}^{(\vec{k}) }   } \nonumber \\
= &{}  \myDoperator{2}{\vec{F}^{(\vec{k})} } - 2 \myDoperator{3}{\vec{F}^{(\vec{k})}}  \nonumber \\
=& {}   \int_{0}^{\infty} (1- F_1^{(k_1)}(x))  (1- F_2^{(k_2)}(x))  \, \mathrm{d}x  \nonumber \\
&{} + \int_{0}^{\infty}  (1- F_2^{(k_2)}(x))  (1- F_3^{(k_3)}(x)) \, \mathrm{d}x \nonumber \\
&{} + \int_{0}^{\infty}  (1- F_3^{(k_3)}(x)) (1- F_1^{(k_1)}(x))  \, \mathrm{d}x \nonumber \\
&{} - 2 \int_{0}^{\infty} (1- F_1^{(k_1)}(x))  (1- F_2^{(k_2)}(x))  (1- F_3^{(k_3)}(x)) \, \mathrm{d}x \nonumber \\
=& \int_{0}^{\infty} \big(  \sum_{n_1 = 0 }^{k_1-1 } e^{- \lambda_1 x} \frac{  (\lambda_1 x)^n_1 }{n_1!}  \big) \big(  \sum_{n_2 = 0 }^{k_2-1 } e^{- \lambda_2 x} \frac{  (\lambda_2 x)^n_2 }{n_2!}  \big)  \, \mathrm{d}x \nonumber  \\
& {} + \int_{0}^{\infty} \big(  \sum_{n_2 = 0 }^{k_2-1 } e^{- \lambda_2 x} \frac{  (\lambda_2 x)^n_2 }{n_2!}  \big)
 \big(  \sum_{n_3 = 0 }^{k_3-1 } e^{- \lambda_3 x} \frac{  (\lambda_3 x)^n_3 }{n_3!}  \big)  \, \mathrm{d}x \nonumber  \\
 &{} + \int_{0}^{\infty} \big(  \sum_{n_1 = 0 }^{k_1-1 } e^{- \lambda_1 x} \frac{  (\lambda_1 x)^n_1 }{n_1!}  \big)
 \big(  \sum_{n_3 = 0 }^{k_3-1 } e^{- \lambda_3 x} \frac{  (\lambda_3 x)^n_3 }{n_3!}  \big)  \, \mathrm{d}x \nonumber  \\
 &{} - 2 \int_{0}^{\infty} \big(  \sum_{n_1 = 0 }^{k_1-1 } e^{- \lambda_1 x} \frac{  (\lambda_1 x)^n_1 }{n_1!}  \big) \big(  \sum_{n_2 = 0 }^{k_2-1 } e^{- \lambda_2 x} \frac{  (\lambda_2 x)^n_2 }{n_2!}  \big) \nonumber \\
 & {} \times  \big(  \sum_{n_3 = 0 }^{k_3-1 } e^{- \lambda_3 x} \frac{  (\lambda_3 x)^n_3 }{n_3!}  \big)  \, \mathrm{d}x \nonumber  \\
= & \frac{1}{\lambda_1+\lambda_2} \sum_{n_1=0}^{k_1-1}\sum_{n_2=0}^{k_2-1}   \frac{  (n_1+n_2)!  }{n_1! n_2!} \big(\frac{\lambda_1}{\lambda_1+\lambda_2}\big)^{n_1}\big(\frac{\lambda_2}{\lambda_1+\lambda_2}\big)^{n_2}  \nonumber \\
& + \frac{1}{\lambda_2+\lambda_3} \sum_{n_2=0}^{k_2-1}\sum_{n_3=0}^{k_3-1} \frac{  (n_2+n_3)!  }{n_2! n_3!} \big(\frac{\lambda_2}{\lambda_2+\lambda_3}\big)^{n_2}\big(\frac{\lambda_3}{\lambda_2+\lambda_3}\big)^{n_3} \nonumber \\
& + \frac{1}{\lambda_3+\lambda_1} \sum_{n_3=0}^{k_3-1}\sum_{n_1=0}^{k_1-1} \frac{  (n_3+n_1)!  }{n_3! n_1!} \big(\frac{\lambda_3}{\lambda_3+\lambda_1}\big)^{n_3} \big(\frac{\lambda_1}{\lambda_3+\lambda_1}\big)^{n_1}  \nonumber \\
& -2  \frac{1}{\lambda_1+\lambda_2+\lambda_3} \sum_{n_1=0}^{k_1-1}\sum_{n_2=0}^{k_2-1} \sum_{n_3=0}^{k_3-1} \frac{  (n_1+n_2+n_3)!  }{n_1! n_2! n_3!} \big(\frac{\lambda_1}{\lambda_1+\lambda_2+\lambda_3}\big)^{n_1} \nonumber \\
& \times   \big(\frac{\lambda_2}{\lambda_1+\lambda_2+\lambda_3}\big)^{n_2}  \big(\frac{\lambda_3}{\lambda_1+\lambda_2+\lambda_3}\big)^{n_3} \eqcomma
\end{align}
using
\begin{align*}
\int_{0}^{\infty }  e^{ax} x^{b-1}\, \mathrm{d}x = \frac{  \Gamma(b)  }{ a^b  } \eqstop
\end{align*}

\end{proof}

%% file: sections/Appendix_C.tex


\section{}
\label{sec:AppendixC}

\subsection{Subgradient}
\begin{myDefinition}[Subgradient]
Let $c: \effectiveDomain{c} \rightarrow \mathbb{R}$  be a real-valued convex function with domain $\effectiveDomain{c}$. For this, a vector $\mathbf{g} \in \mathbb{R}^n$ is a subgradient of $c$ at $\mathbf{x}_0 \in \effectiveDomain{c}$ if $\mathbf{g} \in \partial c(\mathbf{x})$, i.e.
$$c(\mathbf{x})-c(\mathbf{x}_0)\geq \mathbf{g}^T(\mathbf{x}-\mathbf{x}_0), \quad \forall \mathbf{x} \in \effectiveDomain{c} \eqstop$$

\end{myDefinition}

\subsection{Inference for the i.i.d. case}

\begin{proof}[Derivation of   the posterior distribution] 

The posterior is
\begin{align}
p(\lambda | s^{\prime}_{n,1},s^{\prime}_{n,2},\dots,s^{\prime}_{n,j})=\frac{p(s^{\prime}_{n,1},s^{\prime}_{n,2},\dots,s^{\prime}_{n,j}|\lambda) p(\lambda)}{p(s^{\prime}_{n,1},s^{\prime}_{n,2},\dots,s^{\prime}_{n,j})} \eqcomma
\label{eq:posterior_iid_proof}
\end{align}
with the prior distribution
\begin{align*}
p(\lambda)=\Gamma(\lambda|K_0, \lambda_0)=\frac{\lambda_0^{K_0}}{\Gamma(K_0)}\lambda^{K_0-1}e^{-\lambda_0 \lambda}
\end{align*}
and the likelihood
\begin{align*}
p(s^{\prime}_{n,1},s^{\prime}_{n,2},\dots,s^{\prime}_{n,j}|\lambda) = {}&\prod_{i=1}^j \Gamma(s^{\prime}_{n,i} | K_{n.i}, \lambda)   \\
= {}& \prod_{i=1}^j  \frac{\lambda^{K_{n,i}} } {\Gamma(K_{n,i})} s_{n,i}^{\prime K_{n,i}-1} e^{-\lambda s^{\prime}_{n,i}} \eqstop
\end{align*}
Therefore, \eqref{eq:posterior_iid_proof} evaluates to
\begin{align*}
& p(\lambda | s^{\prime}_{n,1},s^{\prime}_{n,2},\dots,s^{\prime}_{n,j})  \\ & =\frac{\frac{\lambda_0^{K_0}}{\Gamma(K_0)}\lambda^{K_0-1}e^{-\lambda_0 \lambda} \prod_{i=1}^j  \frac{\lambda^{K_{n,i}}}{\Gamma(K_{n,i})} s_{n,i}^{\prime K_{n,i}-1}e^{-\lambda s^{\prime}_{n,i}}}{s^{\prime}_{n,1:j}}\\
&=\frac{\lambda_0^{K_0} \prod_{i=1}^j s_{n,i}^{\prime K_{n,i}-1}}{\Gamma(K_0) s^{\prime}_{n,1:j} \prod_{i=1}^j \Gamma(K_{n,i})} \lambda^{K_0-1}e^{-\lambda_0 \lambda}\prod_{i=1}^j \lambda^{K_{n,i}} e^{-\lambda s_{n,i}^{\prime} } \\
& =\frac{1}{Z} \lambda^{K_0-1}\lambda^{\sum_{i=1}^j K_{n,i}} e^{-\lambda_0 \lambda} e^{-\lambda \sum_{i=1}^j s_{n,i}^{\prime}} \eqcomma
\end{align*}
where  $Z$ is a normalization constant.
\begin{align*}
p(\lambda | s^{\prime}_{n,1},s^{\prime}_{n,2},\dots,s^{\prime}_{n,j})&=\frac{1}{Z} \lambda^{(K_0+\sum_{i=1}^j K_{n,i})-1} e^{-(\lambda_0+ \sum_{i=1}^j s_{n,i}^{\prime}) \lambda} \\
&=\Gamma(\lambda| K_{\mathrm{p}},\lambda_{\mathrm{p}}) \eqcomma
\end{align*}
with the posterior parameters
\begin{align*}
	K_{\mathrm{p}}=  {}& K_0+\sum_{i=1}^j K_{n,i}  \eqcomma  \\
\text{and }	\lambda_{\mathrm{p}}=   {} &\lambda_0+\sum_{i=1}^j s^{\prime}_{n,i} \eqstop
\end{align*}

\end{proof}

\subsection{Inference for the Markov modulated case}

\begin{proof}[ML estimates for MMPP] %
	MMPP is a specific example of Hidden Markov Model (HMM) where the observed process $\vec{X}$ is modulated by the (hidden) Markov state  process $\vec{Z}$, as described in Fig.~\ref{fig:Markov_modulated}. Here $\vec{X}|(\vec{Z},\pmb{\theta})$ is exponentially distributed, $\pmb{\theta}$ being the model parameter. In our case, we  observe sum of known number of exponential variables (service times) with same but unknown mean, i.e., Markov modulated gamma variables with known shape but unknown scale parameters. Corresponding (complete) likelihood is given by
	\begin{equation}
	L(\vec{X},\vec{Z}|\pmb{\theta}) = p(z_1|\pmb{\pi})\big{[}\prod_{i=2}^N p(z_i|z_{i-1},\vec{A})\big{]}\prod_{i=1}^N f(x_i|z_i,\pmb{\lambda},m_i)
	\end{equation}
	where $\vec{X}=\{x_1,x_2,..,x_N\}$; $\vec{Z}=\{z_1,z_2,..,z_N\}$; $\pmb{\theta}=\{\boldsymbol{\pi}, \vec{A}, \pmb{\lambda}\}$; $\pmb{\pi}=(\pi_{1},...,\pi_{M})$ and $\vec{A}$ are respectively the  initial distribution and the  transition probability matrix for the hidden states; $\boldsymbol{\lambda}=(\lambda_{1},...,\lambda_{M})$ is the vector of  gamma scale parameters, each corresponding to one of the hidden states; and $m_i's$ are known shape parameters. The integer $M$ denotes number of hidden states and $N$ is number of observations. More specifically, the term corresponding to conditional density of observed states given hidden states means $f(x_i|z_i,\boldsymbol{\lambda},m_i)=f(x_i|\lambda_{l},m_i)$, when $z_i=l$.
	We find ML estimates for $\pmb{\theta}$ using EM algorithm, where at each iteration, we update our estimate for $\pmb{\theta}$ to $\pmb{\theta}^{new}$, the value that maximizes the conditional (complete) log likelihood
\begin{align*}
  Q(\pmb{\theta},\pmb{\theta}^{\text{old} }) = & \sum_{\vec{Z}}p(\vec{Z}|\vec{X},\pmb{\theta}^{\text{old}   } ) \log L(\vec{X},\vec{Z}|\pmb{\theta}) \\
  = & \sum_{k=1}^M \zeta(z_{1k}) \log  \pi_{k}+\sum_{n=2}^N\sum_{j=1}^M\sum_{k=1}^M \xi(z_{n-1,j},z_{nk})  \log \vec{A}_{jk} \\
  &{} + \sum_{n=1}^N\sum_{k=1}^M \zeta(z_{nk})\log f(x_n|\lambda_{k},m_n)  \eqcomma
\end{align*}
	where $\zeta(z_{nk})=P(z_n=k|\vec{X},\pmb{\theta}^{\text{old}})$ and $\xi(z_{l,j},z_{nk})=P(z_l=j,z_n=k|\vec{X},\pmb{\theta}^{\text{old}})$.
The updated estimates can be written as:
\begin{align}
  \pi_{k}   = &{\zeta(z_{1k})}/{\sum_{j=1}^M \zeta(z_{1j})} \eqcomma  \\
  \vec{A}_{jk}  = & {\sum_{n=2}^N\xi(z_{n-1,j},z_{nk})}/{\sum_{l=1}^M \sum_{n=2}^N\xi(z_{n-1,j},z_{nl})} \\
  \lambda_{k} = & {\sum_{n=1}^N\zeta(z_{nk})x_n}/{\sum_{n=1}^N\zeta(z_{nk})m_n} \eqstop
\end{align}


\end{proof}